\documentclass[aps,prl,reprint,showpacs,superscriptaddress,longbibliography,floatfix]{revtex4-2}

\usepackage{mathtools,amssymb,graphicx,units}
\usepackage{amsmath,bm}
\usepackage{bbold}
\usepackage[plainpages=false,pdfpagelabels,colorlinks=true,linkcolor=red,urlcolor=blue,citecolor=blue,pdftitle={Title},pdfauthor={},pdfdisplaydoctitle=true,pdfduplex=DuplexFlipLongEdge]{hyperref}
\usepackage{verbatim}
\usepackage[english]{babel}
\usepackage{xcolor}
\usepackage{dsfont}
\usepackage[normalem]{ulem}

\usepackage{braket}
\usepackage{bbm}

\newcommand{\ketbra}[2]{\mathinner{|{#1}\rangle \! \langle{#2}|}}

\hypersetup{
           breaklinks=true,   % splits links across lines
           colorlinks=true,   % displays links as colored text instead of blocks
           pdfusetitle=true,  % \title and \author values into pdf metadata
        }

\newcommand{\noteSH}[1]{}

\begin{document}

\title{Encoding complex-balanced thermalization in quantum circuits}

\author{Yiting Mao}
\altaffiliation{These authors contributed equally to this work.}
\affiliation{School of Physics, Zhejiang University, Hangzhou, 310058, China}
\affiliation{Beijing Computational Science Research Center, Beijing 100084, China}

\author{Peigeng Zhong}
\altaffiliation{These authors contributed equally to this work.}
\affiliation{School of Physics, Harbin Institute of Technology, Harbin 150001, China}

 \author{Haiqing Lin}
 \email[]{haiqing0@csrc.ac.cn}
 \affiliation{School of Physics, Zhejiang University, Hangzhou, 310058, China}
\affiliation{Institute for Advanced Studies of Physics, Zhejiang University, Hangzhou, 310058, China}

 \author{Xiaoqun Wang}
 \email[]{xiaoqunwang@zju.edu.cn}
 \affiliation{School of Physics, Zhejiang University, Hangzhou, 310058, China}
\affiliation{Institute for Advanced Studies of Physics, Zhejiang University, Hangzhou, 310058, China}

\author{Shijie Hu}
\email[]{shijiehu@csrc.ac.cn}
\affiliation{Beijing Computational Science Research Center, Beijing 100084, China}
\affiliation{Department of Physics, Beijing Normal University, Beijing, 100875, China}

\begin{abstract}
% We propose a protocol for effectively implementing complex-balanced thermalization via Markovian processes on a quantum-circuit platform that couples the system with engineered reservoir qubits.
%\tcb{Thermal reservoirs usually constrain quantum systems through detailed balance, limiting the nonequilibrium steady states that can be reached. 
%Here we show that engineered reservoir qubits provide a programmable microscopic route beyond this constraint by implementing complex-balanced thermalization through Markovian processes.}
Non-Markovian dynamics in open quantum systems often invalidates the complex-balanced thermalization framework, hindering predictive control of quantum simulation platforms designed to prepare out-of-equilibrium states at prescribed temperatures.
We resolve this bottleneck by engineering reservoir qubits as modular microscopic units coupled to a target quantum system and constructing a quantum-circuit platform that enforces strictly Markovian complex-balanced thermalization.
The platform exploits the non-orthogonality of reservoir qubit eigenstates to drive inhomogeneous heating through a modified Kubo-Martin-Schwinger relation, and uses tunable microscopic time-reversibility breaking to generate amplification-dissipation dynamics.
We demonstrate two applications: temporally correlated dichromatic emission and Liouvillian exceptional-point-protected quantum synchronization at finite temperatures, displaying predictive control over out-of-equilibrium state preparation.
%Non-Markovian dynamics in open quantum systems often invalidates the complex-balanced thermalization framework that underpins predictive control of quantum dynamics, posing a major bottleneck for the tailored design of quantum simulation platforms.
%Here, we show that programmable engineered reservoir qubits, coupled as modular microscopic units to a quantum system, can enforce target complex-balanced thermalization via strictly Markovian dynamics, thereby addressing this challenge.
%In this quantum circuit platform, the non-orthogonality of reservoir qubit eigenstates facilitates non-uniform heating through a modified Kubo-Martin-Schwinger relation, while the breaking of microscopic time-reversibility supports amplification-dissipation dynamics, robustly enabling the preparation of out-of-equilibrium states at given temperatures.
%We demonstrate two representative applications of this platform: temporally correlated dichromatic emission and Liouvillian exceptional-point-protected quantum synchronization at finite temperatures, both of which remain experimentally intractable with conventional thermal reservoirs.
%\textcolor{blue}{Our work provides a generalizable framework for programmable non-equilibrium quantum state engineering and scalable quantum simulation with full predictive control over complex-balanced thermalization dynamics.}}
\end{abstract}

\maketitle

High-fidelity quantum manipulation is a prerequisite for scalable quantum computation and quantum simulations~\cite{Georgescu_2014, Eisert_2015, Fauseweh_2024}.
Recent efforts have sought to realize complex balances (CBs) in quantum devices~\cite{Craven2017, Craven2018, Maier2019, Wang_2022, Biehs_2023, Wang2023}, which lead to out-of-equilibrium states (OESs) resembling those observed in studies of kinetic systems~\cite{Horn_1972, Feinberg_1972}, persistent directed flows~\cite{Caloz_2018, Gnesotto2018, Martnez_2019, Yu_2024}, scattering states~\cite{Shi2015}, dissipative synchronizations~\cite{Manzano2013, Koppenhofer2020, Schmolke2022}, and active networks~\cite{Lynn2021, Nartallo_Kaluarachchi_2024, Monti2025, Gladrow2016, Battle2016}.
Despite these advances, the intricate structures of quantum devices and limited microscopic understanding of CB formation have prevented the predictable, on-demand generation of target OESs.

%It is fundamental to note that OESs in quantum devices often arise from the violation of quantum detailed balance (QDB) in microscopically irreversible and non-unitary dynamics~\cite{Denisov_2002, Sanchez_2010, Martnez_2019, Kim_2018, Biehs_2023}, followed by the establishment of CBs~\cite{Craven2017, Craven2018, Maier2019, Wang_2022, Biehs_2023, Wang2023}.
A core mechanism underlying OESs in quantum devices is their typical emergence from the breaking of quantum detailed balance (QDB) via microscopically irreversible and non-unitary dynamics~\cite{Denisov_2002, Sanchez_2010, Martnez_2019, Kim_2018, Biehs_2023}, followed by the establishment of CBs~\cite{Craven2017, Craven2018, Maier2019, Wang_2022, Biehs_2023, Wang2023}.
This process is termed \textit{complex-balanced thermalization} (CBT).
After CBT, the resulting CBs feature dense networks of interconnected transitions across energy levels, enabling richer non-equilibrium dynamics.
By contrast, conventional Boltzmann thermalization~\cite{Breuer_2007, Weiss_2011} enforces a QDB condition characterized by simple pairwise transitions~\cite{Ichiki_2013}.
Critically, the interplay of multiple coupled environments in quantum devices induces non-Markovian effects in CBT, which generally preclude a closed deterministic description of system dynamics based solely on instantaneous transition rates~\cite{Biehs_2023, Craven2017, Craven2018, Maier2019, Wang_2022, Wang2023}.
This constraint not only exposes the limitations of existing CBT dynamics theories~\cite{Wang_2022, Wang2023, Lynn2021, Nartallo_Kaluarachchi_2024, Monti2025, Gladrow2016, Battle2016}, but also hinders the precise control of quantum devices required for OES preparation~\cite{Breuer_2016}.
A Markovian platform for encoding CBT therefore offers a route to resolving this bottleneck by enabling fully traceable microscopic interactions and admitting a solvable set of rate equations that provide a precise description of the platform.
Yet realizing such a platform requires advanced quantum manipulation techniques.

%To resolve this bottleneck, designing a Markovian quantum platform to encode CBT represents a promising avenue,

%In this framework,
%, which goes beyond current phenomenological and data-driven network-reconstruction approaches~\cite{Wang_2022, Wang2023, Lynn2021, Nartallo_Kaluarachchi_2024, Monti2025, Gladrow2016, Battle2016}.

\begin{figure}[b]
\begin{center}
\includegraphics[width=\columnwidth]{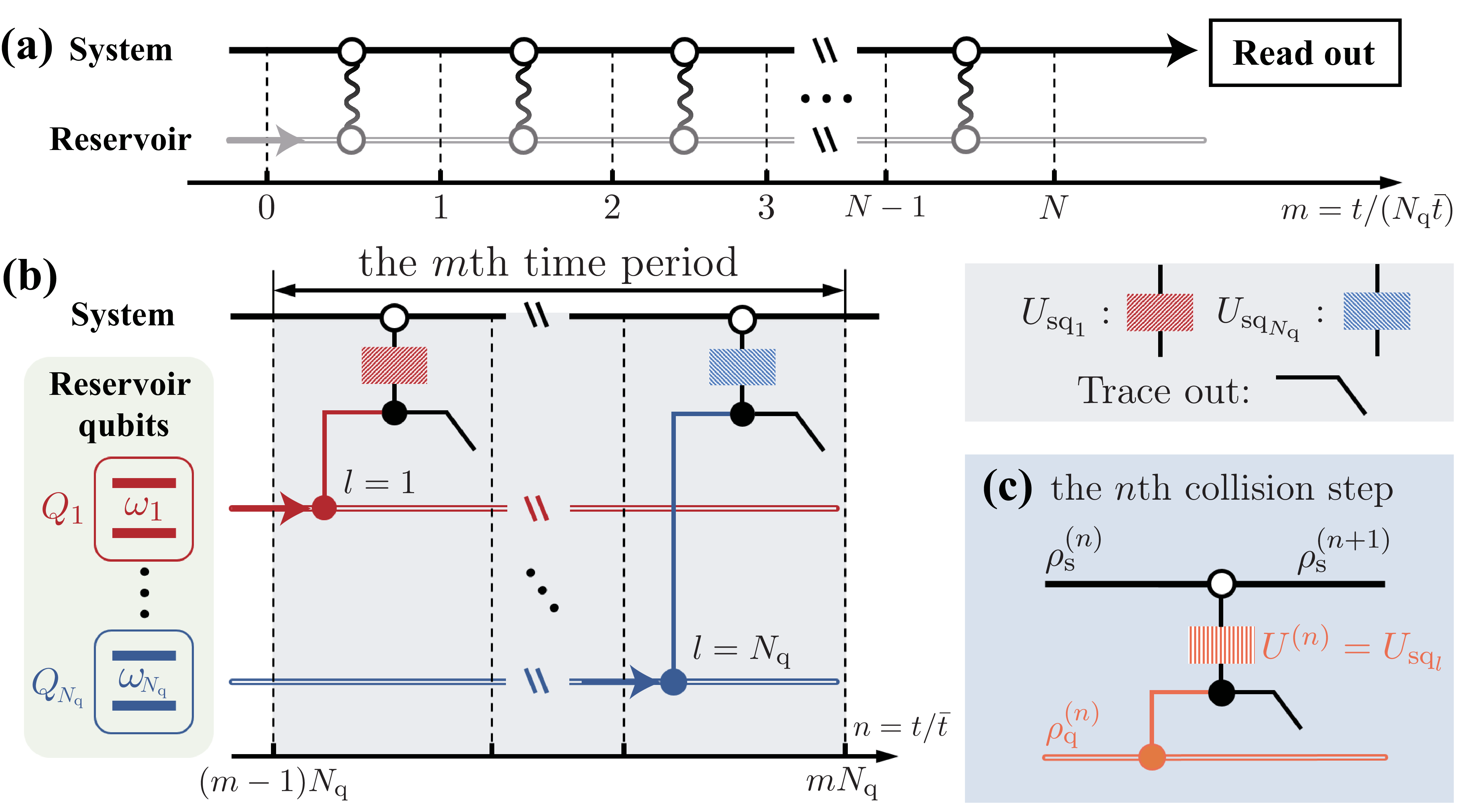}
\caption{\label{fig:fig1} A protocol for quantum circuits. (a) An overview: A quantum system interacts with a qubit set over $N$ time periods.
(b) A period: The system collides in turn with $N_q$ reservoir qubits, labeled $Q_1$, $\cdots$, $Q_{N_\text{q}}$.
(c) Collision step $n = (m-1)N_q + l - 1$: The non-unitary two-qubit gate $U^{(n)}$ couples the system to the qubit q$_l$.
After performing the ``trace out" operation, only the resulting system state participates in subsequent collisions.
}
\end{center}
\end{figure}

Fortunately, recent advancements in qubit control techniques allow for the manipulation of microscopic interactions among multiple qubits with sufficiently long coherence time~\cite{Kjaergaard2020, Krantz2019, Shi2023, Manovitz2025}, which has the potential to be utilized for programming OESs through the design of quantum circuits.
% In this Letter, we present a quantum-circuit protocol for generating these OESs using Markovian processes, implemented by sequentially coupling engineered reservoir qubits to the system and resetting them after each collision.
% The resulting dynamics can be accurately described by a quantum master equation (QME) in the weak-collision limit, ensuring the predictability and precise control of CBT.
In this Letter, we present a protocol for designing reservoir qubits and a well-controllable quantum-circuit platform interfaced with a quantum system.
Our scheme relies solely on Markovian dynamics, yielding a quantum master equation (QME) applicable to the experimentally accessible weak-collision and weak-coupling regime, under which long-term OESs can be robustly predicted.
Consequently, experimentally realizable tuning of reservoir qubit parameters and coupling strengths enables customization of quantum simulations for these OESs.

\textit{All-dissipative-qubit collision realization}. Our quantum circuits [Fig.~\ref{fig:fig1}(a)] couple a system (s) with a set of $N_\text{q}$ non-interacting qubits (q) through a total of $N_\text{c} = N N_\text{q}$ collision steps across $N$ time periods~\cite{Jing_2018, Cusumano_2022, Di_Bartolomeo_2024}.
Each collision lasts a fixed time interval $\bar{t}$.
A time period [Fig.~\ref{fig:fig1}(b)] consists of $N_\text{q}$ collisions, with each collision step involving a single qubit q$_l$ interacting with two energy levels of the system, labeled ``$+$" and ``$-$".
Within a time period, the qubit index $l$ progresses sequentially according to the sequence $\{1$, $\cdots$, $N_\text{q}\}$.
Thus, in the $m$th time period, the system interacts with the qubit q$_l$ at the collision step $n=(m-1)N_\text{q}+l-1$.
Hereafter, we use the collision index $n$ to replace the time-period index $m$ and the qubit index $l$ where appropriate.
While the system involves a Hermitian Hamiltonian $H_\text{s}$, each qubit q$_l$ is described by a non-Hermitian Hamiltonian $H_{\text{q}_l} = \omega_l (\sigma^x_l \cosh \theta_\text{q} + i \sigma^y_l \sinh \theta_\text{q}) / 2$, where $\theta_\text{q} \in [0,\, \pi]$ is an adjustable angle set to be independent of $l$, and $\omega_l > 0$ is the difference of two energy levels for the qubit.
It turns out that the qubit Hamiltonian $H_{\text{q}_l}$ can be readily realized by coupling the qubit $\text{q}_l$ to an external transmon qubit using a well-established post-selection technique~\cite{G_nther_2008, Chen_2021, Zhang_2021, Jebraeilli_2025}.
Meanwhile, $H_{\text{q}_l}$ has energy levels labeled as ``$a$" and ``$b$", corresponding to a real energy spectrum $\{\pm \omega_l / 2\}$ in the $\mathcal{PT}$-unbroken region~\cite{Ruter_2010, Naghiloo_2019}.
These qubits share a common set of biorthonormal left and right eigenstates, denoted by $\bra{a_\text{L}}$, $\bra{b_\text{L}}$, and $\ket{a_\text{R}}$, $\ket{b_\text{R}}$, respectively.
The right eigenstates satisfy the additional convention $\braket{a_\text{R} \vert a_\text{R}} =  \braket{b_\text{R} \vert b_\text{R}} = 1$.

The time evolution of the platform begins at time $t = 0$, corresponding to $n=0$ or equivalently $m=l=1$, where the system is prepared in a state described by the density matrix $\rho^{(0)}_\text{s}$.
Before the $n$th collision [Fig.~\ref{fig:fig1}(c)], the composite system is in a joint state $\rho^{(n)}_\text{s} \otimes \rho^{(n)}_\text{q}$.
The qubit q$_l$ is initialized in the Boltzmann right-eigenstate $\rho^{(n)}_\text{q} = w_a \ketbra{a_\text{R}}{a_\text{R}} + w_b \ketbra{b_\text{R}}{b_\text{R}}$ at an effective preparation temperature $T=1/\beta$~\cite{Du_2022, Roy_2023, Cipolloni_2024, Mao_2024} before being input into the quantum circuits.
The weights are given by $w_{a/b} = e^{\mp \beta \omega_l / 2} / (2 \cosh (\beta \omega_l / 2))$. In practice, $\rho^{(n)}_\text{q}$ corresponds to a coherent state in orthonormal bases and can be efficiently prepared either by a series of single-qubit unitary gates~\cite{Blok_2021, Goss_2022, Luo_2023} or by tracing out an ancillary qubit after a single SU($4$) operation~\cite{Chen_2025}.
At this collision step, one has a bare coupling term $H^{(n)}_{\text{sq}} = A^{(n)} \otimes B^{(n)}$, where operators $A^{(n)}$ and $B^{(n)}$ act on the system and the corresponding qubit q$_l$, respectively.
In particular, $B^{(n)} = \sigma_l^x \cos \theta^{(n)} + \sigma_l^z \sin \theta^{(n)}$ is chosen to account for a mixture of relaxation and dephasing terms by controlling an angle $\theta^{(n)}$~\cite{Poyatos_1996, Diehl_2008, Bacsi_2020, Wang_2020}.
The Hamiltonian for such a composite system can be expressed as $H^{(n)} = H_\text{s} + H^{(n)}_\text{q} + g H^{(n)}_\text{sq}$ with $H^{(n)}_\text{q} \equiv H_{\text{q}_l}$, where $g$ denotes the coupling strength.
Both the two-qubit gate $U^{(n)} = U_{\text{sq}_l} = e^{-i H^{(n)} \bar{t}}$ [shaded rectangle] and the partial ``trace out" of non-orthogonal bases of reservoir qubits $\text{tr}_{\text{q}} [\cdots]$ [black elbow] are feasible in the existing experiments~\cite{Blok_2021, Goss_2022, Luo_2023}.
After the $n$th collision,  $\rho^{(n)}_\text{s}$ evolves into $\rho^{(n+1)}_\text{s}$ with the density matrix
\begin{eqnarray}
\label{eq:CollisionEvolution}
\rho^{(n+1)}_\text{s} = \text{tr}_{\text{q}} \left[ U^{(n)} \left( \rho^{(n)}_\text{s} \otimes \rho^{(n)}_\text{q} \right) U^{(n)\dag} \right]\, .
\end{eqnarray}
% \tcr{Notably, this collision map may not preserve the trace of $\rho^{(n)}_\text{s}$ during time evolution, although it remains Markovian and completely positive.
% Hereafter, we focus on the CBT and the corresponding out-of-equilibrium behavoirs with trace preservation.
The collision map~\eqref{eq:CollisionEvolution} generates Markovian dynamics by erasing reservoir memory through the following two steps.
First, a partial trace over the degrees of freedom of qubit q$_l$ is performed at the end of the $n$th collision to discard system-reservoir correlations.
Second, the qubit q$_l$ is reset to the Boltzmann right-eigenstate $\rho^{(n)}_\text{q}$ prior to the ($n+N_\text{q}$)-th collision in the next time period.
%To characterize CBT, we restrict our discussion to the cases of trace-preserving.
%, where $\text{tr}_\text{s} \rho_\text{s}^{(n)}$ remains constant in the long-term limit $n \rightarrow \infty$.

%\tcb{The weak-collision limit $g\bar{t}\ll 1$} is experimentally feasible in this platform, as demonstrated in the open quantum system simulations using transmon qubits~\cite{GarciaPerez2020IBM, Ciccarello2022Collision}, where $\bar{t}\sim 10$ ns and $g\sim$ 1 MHz give the typical value $g\bar{t}\sim 10^{-2}$.
%\tcb{Further, for weak-coupling $g^2\bar{t}\ll \omega$}, the long-term dynamics are governed exclusively by the resonant part \tcb{$H^{(n)}_\text{sq} \simeq \sum_{\omega= \pm\omega_l} A_\omega^{(n)}\otimes B_{-\omega}^{(n)}$}, which is required to conserve energy in the microscopic subprocesses (see End Matter~A).
%The operators are given by $A_{\pm\omega_l}^{(n)}=\ketbra{\mp}{\mp}\!A^{(n)}\!\ketbra{\pm}{\pm}$, $B_{-\omega_l}^{(n)}=\mathbbm{B}_{ab}^{(n)} \ketbra{a_\text{R}}{b_\text{L}}$ and $B_{\omega_l}^{(n)}=\mathbbm{B}_{ba}^{(n)} \ketbra{b_\text{R}}{a_\text{L}}$, with real coefficients $\mathbbm{B}_{ab}^{(n)} = \braket{a_\text{L} | \! B^{(n)} \! | b_\text{R}}$ and $\mathbbm{B}_{ba}^{(n)} = \braket{b_\text{L} | \! B^{(n)} \! | a_\text{R}}$.

In typical transmon qubits for simulating open quantum systems~\cite{GarciaPerez2020IBM, Ciccarello2022Collision}, with $\bar{t}\sim 10$ ns, $g\sim$ $10$ MHz and $\omega_l = 1$ GHz, we obtain the values $g^2 \bar{t}/\omega_l \sim 10^{-3}$ and $g\bar{t}\sim 0.1$, satisfying both the weak-coupling condition $g^2 \bar{t}\ll \omega_l$ and the weak-collision condition $g\bar{t}\ll 1$.
For the weak coupling $g^2 \bar{t}\ll \omega_l$, the long-term dynamics are governed exclusively by the resonant part $H^{(n)}_\text{sq} \simeq \sum_{\omega= \pm\omega_l} A_\omega^{(n)}\otimes B_{-\omega}^{(n)}$, which is required to conserve energy in the microscopic subprocesses.
The operators are given by $A_{\pm\omega_l}^{(n)}=\ketbra{\mp}{\mp}\!A^{(n)}\!\ketbra{\pm}{\pm}$, $B_{-\omega_l}^{(n)}=\mathbbm{B}_{ab}^{(n)} \ketbra{a_\text{R}}{b_\text{L}}$ and $B_{\omega_l}^{(n)}=\mathbbm{B}_{ba}^{(n)} \ketbra{b_\text{R}}{a_\text{L}}$, with real coefficients $\mathbbm{B}_{ab}^{(n)} = \braket{a_\text{L} | \! B^{(n)} \! | b_\text{R}}$ and $\mathbbm{B}_{ba}^{(n)} = \braket{b_\text{L} | \! B^{(n)} \! | a_\text{R}}$.
Moreover, for the weak collision $g\bar{t}\ll 1$, the difference $\Delta \rho^{(n)}_\text{s} = (\rho^{(n+1)}_\text{s} - \rho^{(n)}_\text{s}) / \bar{t}$ then follows a QME (see End Matter~A)
\begin{eqnarray}\label{eq:GKSL}
\begin{split}
&\Delta \rho^{(n)}_\text{s} = \mathcal{L} [\rho^{(n)}_\text{s}] = -i \left[H_\text{s},\, \rho^{(n)}_\text{s}\right] - \mathcal{L}_\text{d} [\rho^{(n)}_\text{s}] + \mathcal{L}_\text{j} [\rho^{(n)}_\text{s}]\, ,\\
&\mathcal{L}_\text{j} [\rho^{(n)}_\text{s}] \! = \! g^2 \bar{t} \sum_{\omega=\pm \omega_{l}} \bar{\gamma}^{(n)}_{\omega} A^{(n)}_\omega \rho^{(n)}_\text{s} A_\omega^{(n)\dag}\ ,\\
&\mathcal{L}_\text{d} [\rho^{(n)}_\text{s}] \! = \! \frac{g^2 \bar{t}}{2} \sum_{\omega=\pm \omega_{l}} \left\{\gamma^{(n)}_{\omega} A^{(n)}_{-\omega}A^{(n)}_\omega,\,\rho^{(n)}_\text{s} \right\}_\dag\, ,
\end{split}\,
\end{eqnarray}
where $\mathcal{L}_\text{d}$ and $\mathcal{L}_\text{j}$ are the superoperators corresponding to dissipation and quantum jumps, respectively.
Here $\{O_1,O_2\}_{\dagger}\equiv O_1 O_2 + O_2 O^{\dagger}_1$ denotes a generalized anticommutator between operators $O_1$ and $O_2$~\cite{Cao_2023}.
These qubits in the platform, governed by Boltzmann right-eigenstate statistics~\cite{Du_2022, Roy_2023, Cipolloni_2024, Mao_2024}, may drive the system towards achieving CBT, effectively functioning as a specific reservoir that combines the roles of both thermal reservoirs and dissipative sources.
Henceforth, we refer to them as \textit{reservoir qubits}.

In contrast to conventional thermal reservoirs, we need to consider dual spectral functions in this case
\begin{eqnarray}\label{eq:SpectralFunctions}
\begin{split}
\gamma^{(n)}_\omega \!=\!\text{tr}_{\text{q}} \left[\!B_{\omega}^{(n)} B_{-\omega}^{(n)} \rho_\text{q}^{(n)}\!\right]\, ,\ \bar{\gamma}^{(n)}_\omega\!=\!\text{tr}_{\text{q}} \left[\! B_{-\omega}^{(n)\dag} B_{-\omega}^{(n)} \rho^{(n)}_{\text{q}}\!\right].
\end{split}
\end{eqnarray}
Since $\mathbbm{B}_{ab}^{(n)} \ne \mathbbm{B}_{ba}^{(n)*}$, which arises from the non-orthogonality of the right eigenstates $\ket{a_\text{R}}$, $\ket{b_\text{R}}$ of the reservoir qubits in the platform, the two spectral functions $\gamma^{(n)}_\omega$ and $\bar{\gamma}^{(n)}_\omega$ differ (see End Matter B).
\begin{figure}[t]
\begin{center}
\includegraphics[width=\columnwidth]{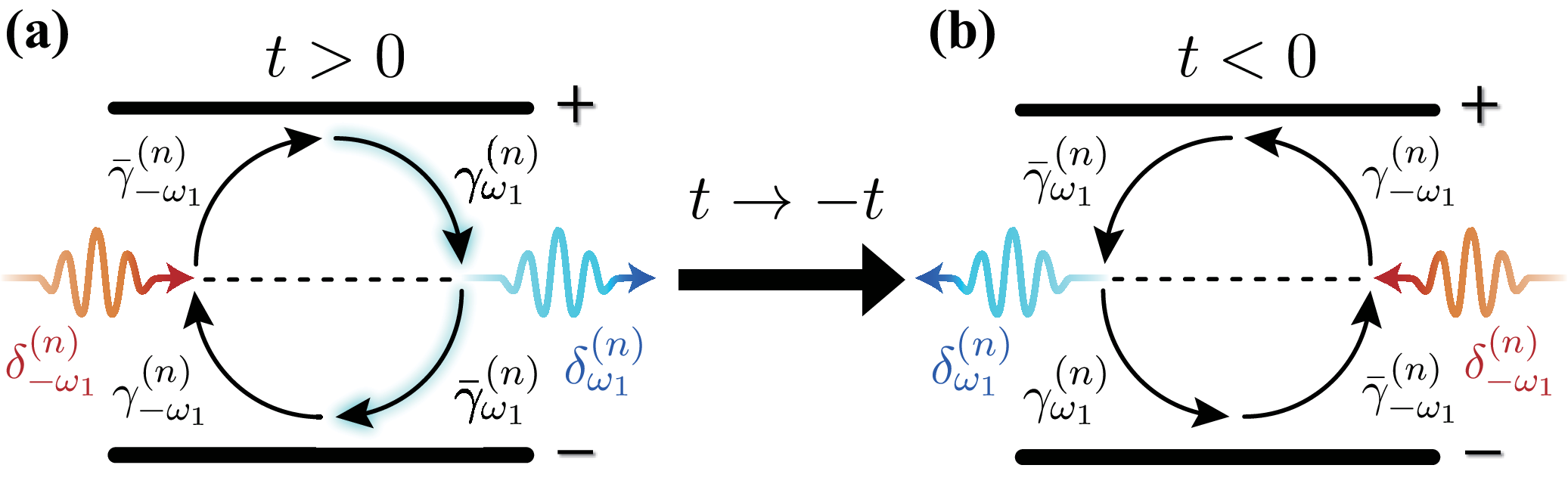}
\caption{\label{fig:fig2} Under time reversal, microscopic subprocesses at (a) positive time $t>0$ and (b) negative time $t<0$.
}
\end{center}
\end{figure}

% reduce
To show the effects of this discrepancy, we consider the simplest scenario in which the system is a qubit interacting with a single reservoir qubit (i.e., $N_\text{q}=1$ and $l=1$).
Eq.~\eqref{eq:GKSL} then simplifies to a Pauli master equation (PME) $\Delta \rho^{(n)}_\text{s} = g^2 \bar{t} \left(\chi^{(n)}_+ \Gamma_{+\rightarrow-} + \chi^{(n)}_- \Gamma_{-\rightarrow+}\right)$, where we only need to consider the diagonal populations $\chi^{(n)}_+ = \braket{+ | \rho^{(n)}_\text{s} | +}$ and $\chi^{(n)}_- =\braket{- | \rho^{(n)}_\text{s} | -} \ge 0$ after thermalization~\cite{FN1}.
The superoperators describing the transition ``$+\rightarrow-$" (from level ``$+$" to level ``$-$") and its reversal ``$-\rightarrow+$" [Fig.~\ref{fig:fig2}(a)] are defined as
\begin{eqnarray}\label{eq:superoperators_positivet}
\begin{split}
\Gamma_{+ \rightarrow -} &= \bar{\gamma}^{(n)}_{\omega_1} \ \ketbra{-}{-} - \gamma^{(n)}_{\omega_1} \ \ketbra{+}{+}\, ,\\
\Gamma_{- \rightarrow +} &= \bar{\gamma}^{(n)}_{-\omega_1} \ketbra{+}{+} - \gamma^{(n)}_{-\omega_1} \ketbra{-}{-}\, ,
\end{split}
\end{eqnarray}
respectively.
A transition, such as ``$+\rightarrow-$" [shaded arrows], contains two microscopic subprocesses: losing probability in level ``$+$" with rate $\gamma^{(n)}_{\omega_1}$ and gaining probability in level ``$-$" with rate $\bar{\gamma}^{(n)}_{\omega_1}$.
% The dynamics in PME so that involve four such subprocesses
The PME dynamics therefore involves four such subprocesses, in contrast to the two pairwise subprocesses in QDB, potentially yielding distinct behaviors.

Under time reversal ($t \rightarrow -t$), the transitions in the PME are reversed: the transition ``$+\rightarrow -$" becomes ``$-\rightarrow +$", accompanied by a change from $+\omega_1$ to $-\omega_1$ and vice versa. This corresponds to swapping $\gamma^{(n)}_{\pm\omega_1}$ and $\bar{\gamma}^{(n)}_{\pm\omega_1}$ in Eq.~\eqref{eq:superoperators_positivet}.
Therefore, for negative time ($t < 0$), the superoperators are given by [Fig.~\ref{fig:fig2}(b)]
\begin{eqnarray}\label{eq:superoperators_negativet}
\begin{split}
\tilde{\Gamma}_{+ \rightarrow -} &= \gamma^{(n)}_{\omega_1} \ \ketbra{-}{-} - \bar{\gamma}^{(n)}_{\omega_1} \ \ketbra{+}{+}\, ,\\
\tilde{\Gamma}_{- \rightarrow +} &= \gamma^{(n)}_{- \omega_1} \ketbra{+}{+} - \bar{\gamma}^{(n)}_{-\omega_1} \ketbra{-}{-}\, ,
\end{split}
\end{eqnarray}
respectively.
Consequently, Eq.~\eqref{eq:superoperators_negativet} cannot be restored from Eq.~\eqref{eq:superoperators_positivet}, meaning that time-reversibility in each subprocess, defined by the conditions of both $\Gamma_{+ \rightarrow -} = \tilde{\Gamma}_{+ \rightarrow -}$ and $\Gamma_{- \rightarrow +} = \tilde{\Gamma}_{- \rightarrow +}$, does not hold~\cite{Roberts_2021}.

Next, the difference between $\gamma^{(n)}_{\pm\omega_1}$ and $\bar{\gamma}^{(n)}_{\pm\omega_1}$, caused by the non-orthogonality of the right eigenstates of the reservoir qubit, introduces dissipation and amplification in the transitions.
For example, in Fig.~\ref{fig:fig2}(a), $\delta^{(n)}_{\omega_1} = \bar{\gamma}^{(n)}_{\omega_1} - \gamma^{(n)}_{\omega_1} < 0$ represents the effective dissipation rate in the transition ``$+\rightarrow -$", while $\delta^{(n)}_{-\omega_1} = \bar{\gamma}^{(n)}_{-\omega_1} - \gamma^{(n)}_{-\omega_1} > 0$ implies the effective amplification rate in its reversal.
% After \tcr{CBT}, dissipation and amplification arising from the platform maintain a vanishing net probability flux
In the long-term limit, the Markovian dynamics governed by QME~\eqref{eq:GKSL} gives rise to CB by balancing dissipation and amplification among the transitions shown in Fig.~\ref{fig:fig2}.
As a direct consequence of CB, the net probability flux vanishes completely
\begin{eqnarray}\label{eq:CB}
\mathcal{J} = \delta^{(n)}_{\omega_1} \chi^{(n)}_+ + \delta^{(n)}_{-\omega_1} \chi^{(n)}_- = 0\, ,
\end{eqnarray}
which corresponds to the eigenmodes of the Liouvillian with zero real parts.

Crucially, the Markovian structure of Eq.~\eqref{eq:CollisionEvolution} is essential for preserving the predictive power of QME for CBs.
If memory effects are not fully eliminated in Eq.~\eqref{eq:CollisionEvolution}, it is generally impossible to derive a closed-form Liouvillian or net probability flux, meaning that computationally expensive time evolution is required to obtain CBs that may be highly sensitive to the choice of initial states.
Even in the rare special cases where closed-form expressions for these quantities can be derived, obtaining long-term solutions for CBs remains a considerable challenge~\cite{Chru_ci_ski_2010, Hegde_2021, Breuer_2016}.

Moreover, the reservoir qubit imposes a constraint \textit{via} a modified Kubo-Martin-Schwinger (KMS) relation
\begin{eqnarray}\label{eq:modifiedKMS}
\eta^{(n)}_{\omega_1} = \bar{\gamma}^{(n)}_{-\omega_1} / \gamma^{(n)}_{\omega_1} = e^{-\beta \omega_1} \mathbbm{B}^{(n)*}_{ba}/\mathbbm{B}^{(n)}_{ab} = e^{-\bar{\beta} \omega_1}
\end{eqnarray}
with the transition-dependent inverse temperature $\bar{\beta} \ne \beta$ when $\mathbbm{B}_{ba}^{(n)\ast}/\mathbbm{B}_{ab}^{(n)}$ is real and positive.
In systems more complex than a single qubit, reservoir qubits ($N_\text{q} > 1$) heat the system non-uniformly, even when a unified $\beta$ is used, which facilitates the establishment of the CBs in CBT.
Below, we present two applications of the platform.

%After CBT, the long-time Liouvillian dynamics~\eqref{eq:GKSL} selects \tcb{the steady state} with the zero Liouvillian eigenvalue, $\mathcal{L}[\rho_\mathrm{s}^{(\infty)}]=0$.
%\tcb{The steady state} is dynamically established by balancing dissipation and amplification in the above transitions, and the zero eigenvalue can be interpreted as a vanishing net probability flux
%\tcb{CB 与 flux 联系起来}

%Notice that the \tcb{Markovian structure} of Eq.~\eqref{eq:CollisionEvolution} is crucial for the prediction of \tcb{CB}. 
%If memory effects were retained in Eq.~\eqref{eq:CollisionEvolution}, \tcb{these transitions} Eq.~(6) would generally depend on \tcb{their memory}, and the \tcb{transient} $\mathcal{J}=0$ would no longer provide a \tcb{closed-form balance condition in general} for the system in the long term.

\begin{figure}[t]
\begin{center}
\includegraphics[width=\columnwidth]{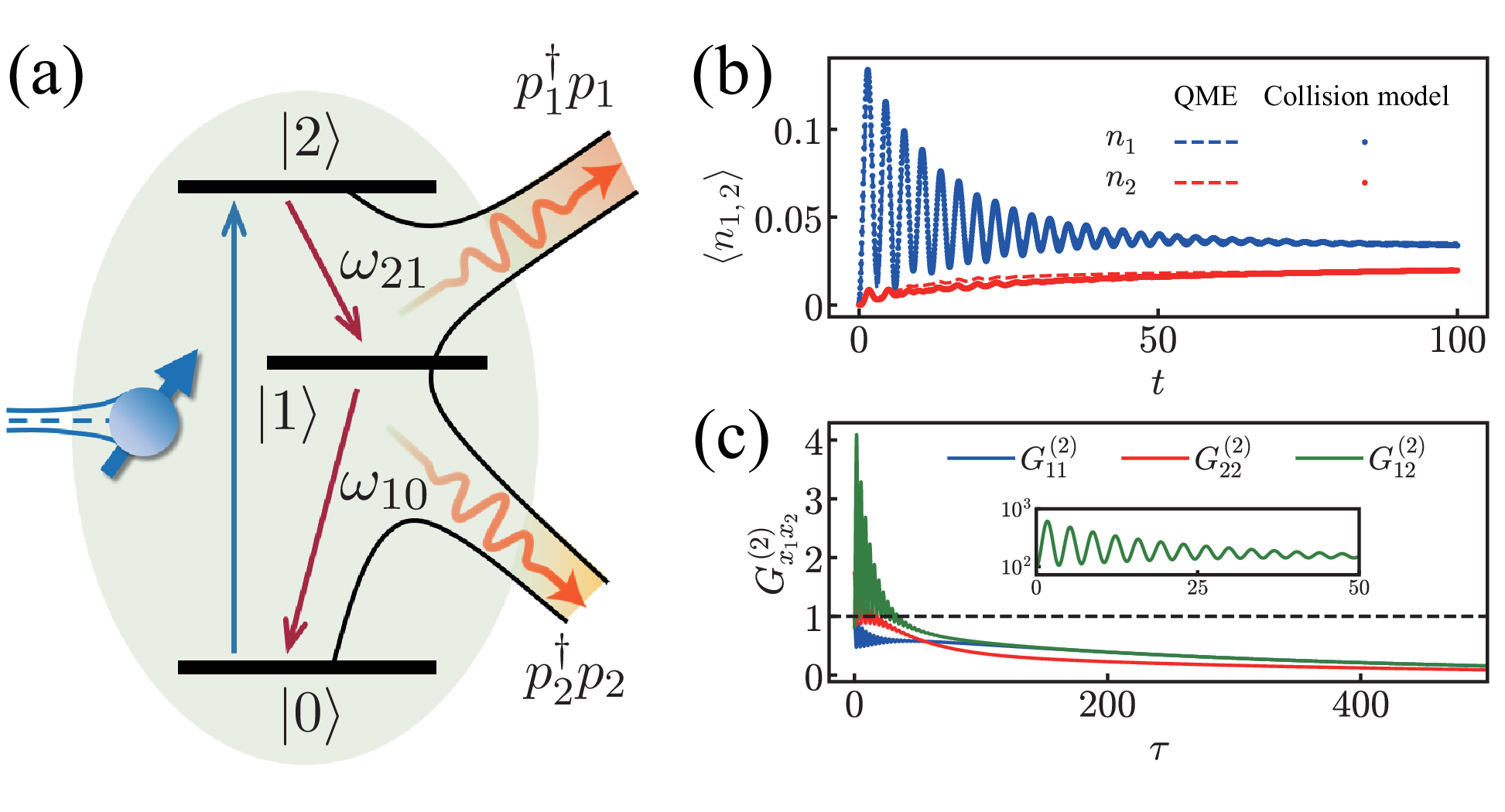}
\caption{\label{fig:fig3}
(a) Dichromatic photon emission setup.
In the setup, a three-level system is coupled to two photonic modes $p_x$ ($x=1$, $2$) through Jaynes-Cummings terms $g_\text{int} (\ketbra{2}{1} p_1 + \ketbra{1}{0} p_2 + \textrm{h.c.} )$.
These photonic modes carry energies $\omega_{21}=0.8$ and $\omega_{10}=1$, respectively, without detuning.
Photon emission is modeled using additional Lindblad operators $L_x = \sqrt{\kappa} p_x$, as detailed in SM~\cite{SM}.
The three-level system interacts with $N_\text{q}=3$ qubits through the quantum-circuit platform [Fig.~\ref{fig:fig1}].
(b) Time-evolving photon numbers $\braket{n_{1/2}}$.
(c) Second-order time correlation functions $G_{x_1 x_2}^{(2)}$ for $t=+\infty$.
Inset: data for small $\tau$ with $\theta_\text{q} \approx \theta_\text{q}^\text{LEP}$ (LEP $\cosh \theta_\text{q}^\text{LEP}=2$).
We used spin-$1$ operator $A^{(n)}=\mathcal{S}^x$, and parameters $\theta^{(n)}=\pi/3$, $\beta=1$, $g=1$, $\bar{t}=0.05$, $g_\text{int}=0.4$ and $\kappa=0.1$.
For main panels (b, c), $\theta_\text{q}=\pi/6$.
}
\end{center}
\end{figure}

% \textit{Temporally correlated dichromatic emission}.---This platform can induce strong temporal correlations in light emission via CBs.
%\textit{Temporally correlated dichromatic emission}.---\tcr{Via CBs, this platform can induce strong time-ordered correlations between two-color emissions from different transitions of the same emitter, which are difficult to access with Hermitian thermal reservoirs.}
\textit{Temporally correlated dichromatic emission}.---On this platform, the establishment of CB enables the activation of two-color photon emission in a specific time-ordered sequence.
In Fig.~\ref{fig:fig3}(a), we consider a three-level system with energy levels $\ket{0}$, $\ket{1}$, $\ket{2}$, coupled to two photonic modes $p_1$ and $p_2$, and denote their joint density matrix by $\rho_\text{sp}$.
When $g\bar{t} \ll 1$ and $g^2 \bar{t} \ll \omega_{21}$, $\omega_{10}$, satisfying the weak-collision and weak-coupling conditions, the photon numbers $\braket{n_1}$ and $\braket{n_2}$ obtained from the collision map in Eq.~\eqref{eq:CollisionEvolution} and the QME treatment described in Supplemental Material (SM)~\cite{SM} are in excellent agreement, quantitatively describing time evolution towards nonequilibrium steady states in the long term [Fig.~\ref{fig:fig3}(b)].

During time evolution, we monitor the second-order time-correlation function~\cite{Brown1956, Glauber1963}
\begin{eqnarray}\label{eq:2ndordercorrelationfunc}
G_{x_1 x_2}^{(2)} (n^\prime) =\frac{\braket{p^\dag_{x_1} p_{x_2}^{(n^\prime)\dag} p_{x_2}^{(n^\prime)} p_{x_1}}_n} {\braket{p^\dag_{x_1} p_{x_1}}_n \braket{p^\dag_{x_2} p_{x_2}}_n}
\end{eqnarray}
to explore the properties of this dichromatic light.
% Here, the expectation value $\braket{\cdots}_n = \text{tr}_\text{s} [\rho^{(n)}_\text{s} \cdots] / \text{tr}_\text{s} \rho^{(n)}_\text{s}$ is measured immediately after the $n$th collision ($t = n \bar{t}$), and $p_{x_2}^{(n^\prime)}$ denotes the annihilation operator delayed by $n^\prime$ time steps, with a time lag $\tau = n^\prime \bar{t}$.
Here, the expectation value $\braket{\cdots}_n = \text{tr}_\text{sp} [\rho^{(n)}_\text{sp} \cdots] / \text{tr}_\text{sp} \rho^{(n)}_\text{sp}$ is measured immediately after the $n$th collision ($t = n \bar{t}$), and $p_{x_2}^{(n^\prime)}$ denotes the annihilation operator delayed by $n^\prime$ time steps, with a time lag $\tau = n^\prime \bar{t}$.
Using conventional thermal reservoirs, the photon field is expected to exhibit thermal bunching, with self correlations $G_{11}^{(2)} = G_{22}^{(2)} \approx 2$ and the cross correlation $G_{12}^{(2)} \approx 1$ at $\tau = 0$, all of which decay exponentially to $1$ within the memory time of the system~\cite{Glauber1963}.
In contrast, when driven by the reservoir qubits in our platform, the emitted photons show enhanced photon bunching, with $G_{12}^{(2)} \gg 2$ over a narrow region of small $\tau$.
At longer delays, the cross correlations become suppressed, with $G_{12}^{(2)}<1$ at large $\tau$ [Fig.~\ref{fig:fig3}(c)].
This short-time-lag (STL) enhancement of photon bunching and long-time-lag (LTL) suppression of the photon (pair) emission originates from the strong temporal correlation between transitions $2 \rightarrow 1$ and $1\rightarrow 0$ [red arrows] involved in the established CBs [Fig.~\ref{fig:fig3}(a)].
The system emits photons in a rapid cascade: $2 \rightarrow 1 \rightarrow 0$, producing strong STL correlations.
The mode is then depleted until it is recharged by thermal pumping $0 \rightarrow 2$ [blue arrow], which suppresses the LTL photon emission.
We also find that the STL bunching for small $\tau$ grows rapidly as $\theta_\text{q}$ gets close to the Liouvillian exceptional point (LEP) [Fig.~\ref{fig:fig3}(c) inset].

These enhanced temporal correlations differ from the thermal correlations in regular thermal light~\cite{Gatti2004, Valencia2005, Zhang2005, Tan2023, Lee2023} or the quantum correlations enhanced by complex nonlinear photonic processes~\cite{Burnham_1970, Couteau_2018}.
With fine tuning, the generated strongly temporally correlated dichromatic photonic modes facilitate the realization of relevant correlation-based sensing techniques, such as ghost imaging~\cite{Gatti2004, Valencia2005, Zhang2005} and two-photon lidar~\cite{Tan2023, Lee2023}.

\begin{figure}[t]
\begin{center}
\includegraphics[width=\columnwidth]{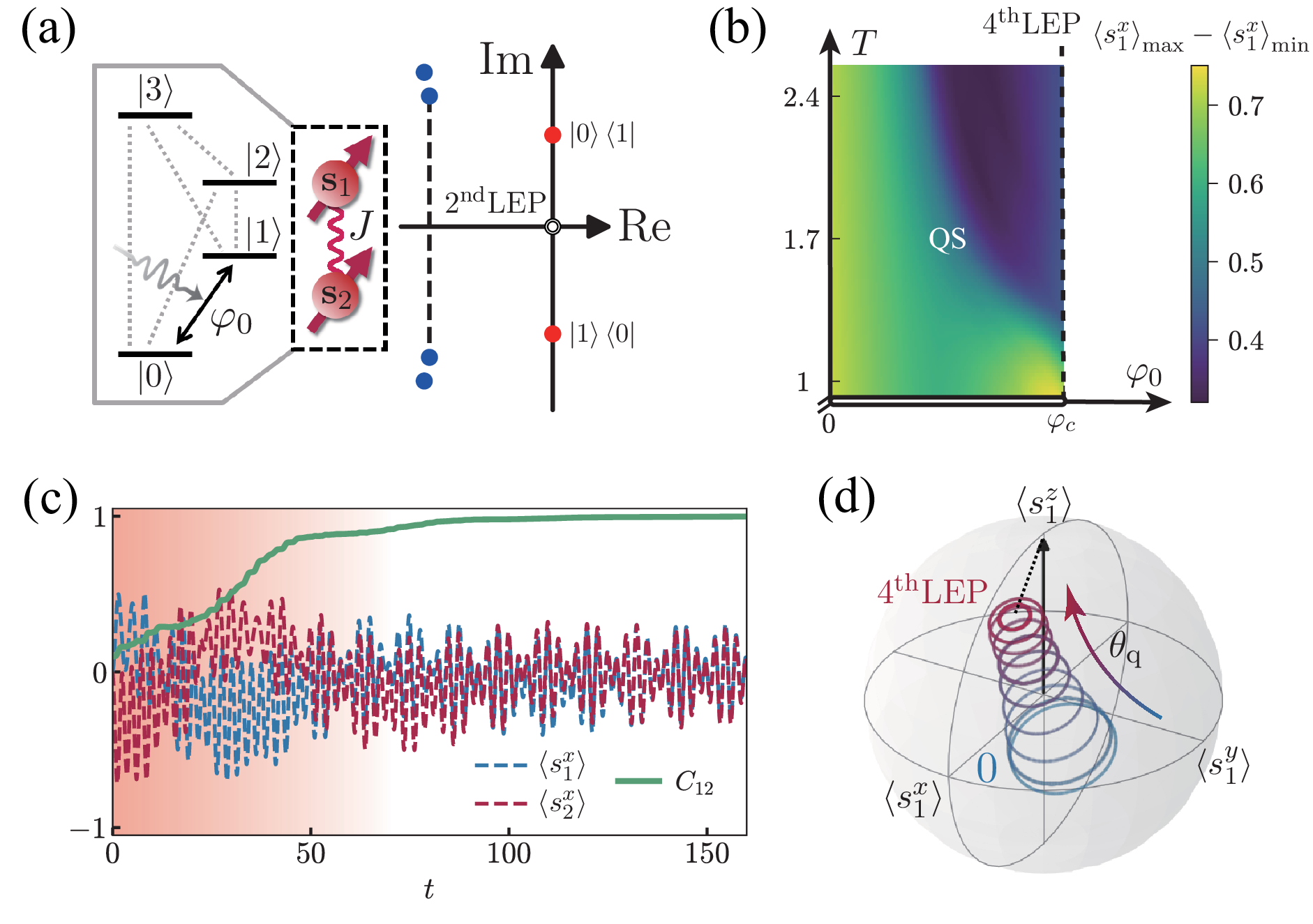}
\caption{\label{fig:fig4}
(a) A LEP-protected quantum synchronization setup.
In the setup, two spins $\mathbf{s}_1$ and $\mathbf{s}_2$ (momenta $|\mathbf{s}_1|=|\mathbf{s}_2|=1/2$) are coupled with an Ising-type interaction and modulated by external magnetic fields along both the $x$ and $z$ axes.
The Hamiltonian $H_\text{s} = J s^z_1 s^z_2 + h_x (s^x_1 + s^x_2) + h_z (s^z_1 + s^z_2)$ provides a spectrum of four energy levels: $\ket{0}$ (ground state), $\ket{1}$ (first excited state), $\ket{2}$ and $\ket{3}$, where $J$, $h_x$ and $h_z$ are the strengths of the interaction and fields.
These two spins interact with $6$ reservoir qubits through the quantum-circuit platform [Fig.~\ref{fig:fig1}].
(b) Phase diagram when $\theta_\text{q}=0.55$. The QS region is shaded according to the long-term oscillation amplitude of $\braket{s^x_1}$.
(c) Time-evolving $\braket{s^x_1}$, $\braket{s^x_2}$ and Pearson coefficient $C_{12}$ obtained from the collision map with $\theta_\text{q}=0.55$ and $\varphi_0=\pi/3$.
(d) Time-evolution trajectory of spin $\braket{\mathbf{s}_1}=(\braket{s^x_1},\braket{s^y_1},\braket{s^z_1})$ on the Bloch sphere for different $\theta_\text{q}$ with $\varphi_0=\pi/3$ fixed.
At the fourth-order LEP $\theta_\text{q} = \text{arctanh}\left(\sin\varphi_\text{c}\right)\approx 1.317$, the system can no longer sustain balanced amplification-dissipation driving.
We choose $A^{(n)}=s^x_1$, $J = 0.2$, $h_z = 2 h_x = 1$, $\beta=1$, $g=2$ and $\bar{t}=0.05$.
}
\end{center}
\end{figure}

\textit{LEP-protected quantum synchronization at finite temperatures}.---Likewise, on this platform, $N_\text{q}=6$ reservoir qubits perfectly lock the stable phase coherence evolution between two spins [Fig.~\ref{fig:fig4}(a) left], thereby activating their long-term quantum synchronization (QS).
%\tcr{Under detailed balance, thermal fluctuations tend to wash out phase coherence and degrade quantum synchronization between the two spins.
%\tcr{By contrast, breaking the detailed balance permits a LEP-protected oscillation mode that preserves synchronization even at finite temperatures}}.
For simplicity, we modulate only the coupling angle $\theta^{(n)} = \varphi_0$ when the collision is associated with the transition ``$0 \leftrightarrow 1$" between the ground state $\ket{0}$ and the first excited state $\ket{1}$.
For the other transitions, we fix $\sin\theta^{(n)} = \sin \varphi_\text{c} = \tanh \theta_\text{q}$.
This platform enables QS over a wide parameter region where $\varphi_0 < \varphi_\text{c}$ and temperature $T > 0$ is finite [colored region in Fig.~\ref{fig:fig4}(b)].
% In this region, two Liouvillian eigenstates with zero eigenvalues coalesce, resulting in a rank-$2$ LEP.
% One of the coalescing eigenstates originates from the system Hamiltonian, while the other is induced by the modified KMS relation.
% The pair of coalescing eigenstates is associated with a pair of \textit{oscillation modes}, e.g. $\ketbra{0}{1}$ and $\ketbra{1}{0}$, which have the maximal real part and form a conjugate pair along the imaginary axis of the Liouvillian spectrum [Fig.~\ref{fig:fig4}(a) right].
% These oscillation modes govern the long-term QS dynamics, thus protecting QS through the LEP.
%\tcr{In this region, the Liouvillian contains two zero-eigenvalue modes and a conjugate pair of oscillation modes with zero real parts.
%The zero-eigenvalue modes support the trace-preserving population dynamics, with one mode inherited from one of Hamiltonian eigenstates and the other induced by the modified KMS relation, resulting in a rank-$2$ LEP.
%The oscillation modes, associated with coherences such as $\ketbra{0}{1}$ and $\ketbra{1}{0}$, lie on the imaginary axis [Fig.~\ref{fig:fig4}(a) right], with their oscillation frequency equal to $\omega_{10}$, 
%While the modified KMS relation tunes their real parts to zero, thereby protecting the long-term QS dynamics.}
In this region, the long-term dynamics is governed by two zero-eigenvalue modes and a conjugate pair of oscillation modes with zero real parts, all of which lie on the imaginary axis of the Liouvillian spectrum [Fig.~\ref{fig:fig4}(a) right], as enforced by the modified KMS relation~\eqref{eq:modifiedKMS}.
The zero-eigenvalue modes form a rank-$2$ LEP, sharing one of the Hamiltonian eigenstates $\ket{0}$, while the oscillation modes are associated with the off-diagonal parts $\ketbra{0}{1}$ and $\ketbra{1}{0}$ in $\rho^{(n)}_\text{s}$, sustaining the evolution of phase coherence.
Upon violation of the LEP due to the breakdown of Eq.~\eqref{eq:modifiedKMS} for $\ket{0}$, the eigenvalues associated with the oscillation modes exhibit negative real parts. These modes decay in the long term, leading to the loss of QS.
Thus, the long-term QS dynamics is protected by the LEP.
At zero preparation temperature, the QS of two spins vanishes in the absence of thermal fluctuations.

To quantify QS, we compute the Pearson correlation
\begin{eqnarray}
C_{12}(t)=\frac{\overline{(s_1^x - \overline{s_1^x}) (s_2^x - \overline{s_2^x})}}{\sqrt{\overline{(s^x_1-\overline{s^x_1})^2}\ \overline{(s^x_2-\overline{s^x_2})^2}}}\, ,
\end{eqnarray}
where the expectation value $\overline{\mathcal{O}} =  (1/n') \sum^{n+n'-1}_{z=n} \braket{\mathcal{O}}_z$ is measured at $t = n \bar{t}$ and then averaged over $n' = 2000$ collision steps.
Starting from the initial state $\rho_\text{s}^{(0)} = \ketbra{\uparrow_1 \downarrow_2}{\uparrow_1 \downarrow_2}$, which represents the antiparallel configuration for the two spins, $C_{12}$ evolves towards $1$ in the long term.
This indicates perfect in-phase QS [Fig.~\ref{fig:fig4}(c)], which is protected by LEP.
In the case of a ferromagnetic Ising-type interaction, the platform yields perfect anti-phase QS with $C_{12} = -1$ (see SM~\cite{SM}).

With $\varphi_0$ fixed, we plot the envelopes (nearly a circle) of the time-evolution trajectory of spin $\braket{\textbf{s}_1}$ on the Bloch sphere [Fig.~\ref{fig:fig4}(d)].
This demonstrates that the platform allows continuous control over the accessible QS states, because $\theta_\text{q}$ resets the thermal excitation rate at each transition, as indicated by the modified KMS relation in Eq.~\eqref{eq:modifiedKMS}, thereby controlling the coherence between $\ket{0}$ and $\ket{1}$.
These envelopes form a cone, with the vertex corresponding to the fourth-order LEP at $\varphi_0 = \varphi_\text{c}$.
%To align the collision map with QME, an additional von-Hove approximation $g\rightarrow0$ must also be considered~\cite{SM}.
%\tcr{
%Since the first Liouvillian gap gets closure in the QS regime, the oscillation modes are not controlled solely by $g\bar{t}$, and decreasing $\bar{t}$ at fixed $g$ and reducing $g$ at fixed $\bar{t}$ show distinct convergence behaviors (see SM for details).
%To align the collision map with the QME dynamics here, an additional von-Hove weak-coupling condition $g\rightarrow 0$ must also be considered.
%}
%Systematic tests for the optimal choice of $\rho_\text{s}^{(0)}$, $g$ and $\bar{t}$ are provided in SM~\cite{SM}.
More tests for distinct $\rho_\text{s}^{(0)}$ are provided in SM~\cite{SM}.

\textit{Summary and discussion}.---We have proposed a protocol to simulate complex-balanced thermalization on a quantum-circuit platform that couples the system with engineered reservoir qubits.
These qubits serve two functions: they act as thermal reservoirs that assign Boltzmann weight distributions to differentiate between high- and low-energy levels, and generate dissipation due to the non-orthogonality of eigenstate wave functions. 
This platform can reach the weak-collision and weak-coupling regime and enable well-controlled complex-balanced dynamics.
%produces non-uniform heating via a modified Kubo-Martin-Schwinger relation,
Using this platform, we demonstrate two applications.
Moreover, our protocol makes the violation of detailed balance programmable through engineered reservoir-qubit non-orthogonality and coupling operators on the quantum-circuit platform, extending beyond recent advances in synthetic fields~\cite{Alicki_2023, Biehs_2023}.
%In contrast to recent works where detailed balance violation emerges in open quantum systems beyond the Born approximation or in synthetic fields~\cite{Alicki_2023, Biehs_2023}, our protocol makes the violation programmable in the quantum-circuit platform.

The dual spectral functions in quantum master equations will help connect our work to non-Hermitian quantum fluctuation relations and non-Hermitian linear response theory~\cite{Pan2020, Geier2022}, potentially enabling applications based on non-unitary time evolution~\cite{terashima2005, Harrow2009, Uola2020}.
Moreover, extending our protocol to treat situations beyond those considered in the current work, e.g., non-trace-preserving dynamics, is an important direction for  further work.
%\textcolor{red}{The extension of dual spectral functions may potentially establish connections between our work and non-Hermitian quantum fluctuation relations as well as non-Hermitian linear response theory~\cite{Pan2020, Geier2022}, thereby inspiring a wide range of applications based on non-unitary time evolution~\cite{terashima2005, Harrow2009, Uola2020}.
%Moreover, cases without trace preservation merits further exploration.}
%Although the collision map in Eq.~\eqref{eq:CollisionEvolution} may fail to be trace-preserving, meaning that the quantum master equation in Eq.~\eqref{eq:GKSL} does not necessarily generate a completely positive, trace-preserving quantum dynamical semigroup~\cite{Lindblad_1976}, it is noteworthy that the Markovian dynamics of the system can still be effectively captured by Eq.~\eqref{eq:GKSL}.
%This suggests that a more comprehensive theory, involving dual spectral functions, merits further exploration.
%Moreover, this investigation could establish connections between our work, non-Hermitian quantum fluctuation relations, and non-Hermitian linear response theory~\cite{Pan2020, Geier2022}, potentially inspiring a wide range of applications based on non-unitary time evolution~\cite{terashima2005, Harrow2009, Uola2020}.

\begin{acknowledgments}
We are grateful to Yihan Yu for fruitful discussions.
This work is supported by the MOST (Grants No.~2022YFA1402700), the NSFC (Grants No.~12174020 and 12574163), and the FRFCU (Grant No.~AUGA5710025425).
Computational resources from Tianhe-2JK at the Beijing Computational Science Research Center and Quantum Many-body-{\rm I} cluster at SPA, Shanghai JiaoTong University are also highly appreciated.
\end{acknowledgments}

\bibliography{refs}

\begin{thebibliography}{79}%
\makeatletter
\providecommand \@ifxundefined [1]{%
 \@ifx{#1\undefined}
}%
\providecommand \@ifnum [1]{%
 \ifnum #1\expandafter \@firstoftwo
 \else \expandafter \@secondoftwo
 \fi
}%
\providecommand \@ifx [1]{%
 \ifx #1\expandafter \@firstoftwo
 \else \expandafter \@secondoftwo
 \fi
}%
\providecommand \natexlab [1]{#1}%
\providecommand \enquote  [1]{``#1''}%
\providecommand \bibnamefont  [1]{#1}%
\providecommand \bibfnamefont [1]{#1}%
\providecommand \citenamefont [1]{#1}%
\providecommand \href@noop [0]{\@secondoftwo}%
\providecommand \href [0]{\begingroup \@sanitize@url \@href}%
\providecommand \@href[1]{\@@startlink{#1}\@@href}%
\providecommand \@@href[1]{\endgroup#1\@@endlink}%
\providecommand \@sanitize@url [0]{\catcode `\\12\catcode `\$12\catcode
  `\&12\catcode `\#12\catcode `\^12\catcode `\_12\catcode `\%12\relax}%
\providecommand \@@startlink[1]{}%
\providecommand \@@endlink[0]{}%
\providecommand \url  [0]{\begingroup\@sanitize@url \@url }%
\providecommand \@url [1]{\endgroup\@href {#1}{\urlprefix }}%
\providecommand \urlprefix  [0]{URL }%
\providecommand \Eprint [0]{\href }%
\providecommand \doibase [0]{https://doi.org/}%
\providecommand \selectlanguage [0]{\@gobble}%
\providecommand \bibinfo  [0]{\@secondoftwo}%
\providecommand \bibfield  [0]{\@secondoftwo}%
\providecommand \translation [1]{[#1]}%
\providecommand \BibitemOpen [0]{}%
\providecommand \bibitemStop [0]{}%
\providecommand \bibitemNoStop [0]{.\EOS\space}%
\providecommand \EOS [0]{\spacefactor3000\relax}%
\providecommand \BibitemShut  [1]{\csname bibitem#1\endcsname}%
\let\auto@bib@innerbib\@empty
%</preamble>
\bibitem [{\citenamefont {Georgescu}\ \emph {et~al.}(2014)\citenamefont
  {Georgescu}, \citenamefont {Ashhab},\ and\ \citenamefont
  {Nori}}]{Georgescu_2014}%
  \BibitemOpen
  \bibfield  {author} {\bibinfo {author} {\bibfnamefont {I.}~\bibnamefont
  {Georgescu}}, \bibinfo {author} {\bibfnamefont {S.}~\bibnamefont {Ashhab}},\
  and\ \bibinfo {author} {\bibfnamefont {F.}~\bibnamefont {Nori}},\ }\bibfield
  {title} {\bibinfo {title} {Quantum simulation},\ }\href
  {https://doi.org/10.1103/revmodphys.86.153} {\bibfield  {journal} {\bibinfo
  {journal} {Reviews of Modern Physics}\ }\textbf {\bibinfo {volume} {86}},\
  \bibinfo {pages} {153} (\bibinfo {year} {2014})}\BibitemShut {NoStop}%
\bibitem [{\citenamefont {Eisert}\ \emph {et~al.}(2015)\citenamefont {Eisert},
  \citenamefont {Friesdorf},\ and\ \citenamefont {Gogolin}}]{Eisert_2015}%
  \BibitemOpen
  \bibfield  {author} {\bibinfo {author} {\bibfnamefont {J.}~\bibnamefont
  {Eisert}}, \bibinfo {author} {\bibfnamefont {M.}~\bibnamefont {Friesdorf}},\
  and\ \bibinfo {author} {\bibfnamefont {C.}~\bibnamefont {Gogolin}},\
  }\bibfield  {title} {\bibinfo {title} {Quantum many-body systems out of
  equilibrium},\ }\href {https://doi.org/10.1038/nphys3215} {\bibfield
  {journal} {\bibinfo  {journal} {Nature Physics}\ }\textbf {\bibinfo {volume}
  {11}},\ \bibinfo {pages} {124} (\bibinfo {year} {2015})}\BibitemShut
  {NoStop}%
\bibitem [{\citenamefont {Fauseweh}(2024)}]{Fauseweh_2024}%
  \BibitemOpen
  \bibfield  {author} {\bibinfo {author} {\bibfnamefont {B.}~\bibnamefont
  {Fauseweh}},\ }\bibfield  {title} {\bibinfo {title} {Quantum many-body
  simulations on digital quantum computers: State-of-the-art and future
  challenges},\ }\href {https://doi.org/10.1038/s41467-024-46402-9} {\bibfield
  {journal} {\bibinfo  {journal} {Nature Communications}\ }\textbf {\bibinfo
  {volume} {15}},\ \bibinfo {pages} {2123} (\bibinfo {year}
  {2024})}\BibitemShut {NoStop}%
\bibitem [{\citenamefont {Craven}\ and\ \citenamefont
  {Nitzan}(2017)}]{Craven2017}%
  \BibitemOpen
  \bibfield  {author} {\bibinfo {author} {\bibfnamefont {G.~T.}\ \bibnamefont
  {Craven}}\ and\ \bibinfo {author} {\bibfnamefont {A.}~\bibnamefont
  {Nitzan}},\ }\bibfield  {title} {\bibinfo {title} {{Electrothermal Transistor
  Effect and Cyclic Electronic Currents in Multithermal Charge Transfer
  Networks}},\ }\href {https://doi.org/10.1103/PhysRevLett.118.207201}
  {\bibfield  {journal} {\bibinfo  {journal} {Physical Review Letters}\
  }\textbf {\bibinfo {volume} {118}},\ \bibinfo {pages} {207201} (\bibinfo
  {year} {2017})}\BibitemShut {NoStop}%
\bibitem [{\citenamefont {Craven}\ \emph {et~al.}(2018)\citenamefont {Craven},
  \citenamefont {He},\ and\ \citenamefont {Nitzan}}]{Craven2018}%
  \BibitemOpen
  \bibfield  {author} {\bibinfo {author} {\bibfnamefont {G.~T.}\ \bibnamefont
  {Craven}}, \bibinfo {author} {\bibfnamefont {D.}~\bibnamefont {He}},\ and\
  \bibinfo {author} {\bibfnamefont {A.}~\bibnamefont {Nitzan}},\ }\bibfield
  {title} {\bibinfo {title} {{Electron-Transfer-Induced Thermal and
  Thermoelectric Rectification}},\ }\href
  {https://doi.org/10.1103/PhysRevLett.121.247704} {\bibfield  {journal}
  {\bibinfo  {journal} {Physical Review Letters}\ }\textbf {\bibinfo {volume}
  {121}},\ \bibinfo {pages} {247704} (\bibinfo {year} {2018})}\BibitemShut
  {NoStop}%
\bibitem [{\citenamefont {Maier}\ \emph {et~al.}(2019)\citenamefont {Maier},
  \citenamefont {Brydges}, \citenamefont {Jurcevic}, \citenamefont {Trautmann},
  \citenamefont {Hempel}, \citenamefont {Lanyon}, \citenamefont {Hauke},
  \citenamefont {Blatt},\ and\ \citenamefont {Roos}}]{Maier2019}%
  \BibitemOpen
  \bibfield  {author} {\bibinfo {author} {\bibfnamefont {C.}~\bibnamefont
  {Maier}}, \bibinfo {author} {\bibfnamefont {T.}~\bibnamefont {Brydges}},
  \bibinfo {author} {\bibfnamefont {P.}~\bibnamefont {Jurcevic}}, \bibinfo
  {author} {\bibfnamefont {N.}~\bibnamefont {Trautmann}}, \bibinfo {author}
  {\bibfnamefont {C.}~\bibnamefont {Hempel}}, \bibinfo {author} {\bibfnamefont
  {B.~P.}\ \bibnamefont {Lanyon}}, \bibinfo {author} {\bibfnamefont
  {P.}~\bibnamefont {Hauke}}, \bibinfo {author} {\bibfnamefont
  {R.}~\bibnamefont {Blatt}},\ and\ \bibinfo {author} {\bibfnamefont {C.~F.}\
  \bibnamefont {Roos}},\ }\bibfield  {title} {\bibinfo {title}
  {{Environment-Assisted Quantum Transport in a 10-qubit Network}},\ }\href
  {https://doi.org/10.1103/PhysRevLett.122.050501} {\bibfield  {journal}
  {\bibinfo  {journal} {Physical Review Letters}\ }\textbf {\bibinfo {volume}
  {122}},\ \bibinfo {pages} {050501} (\bibinfo {year} {2019})}\BibitemShut
  {NoStop}%
\bibitem [{\citenamefont {Wang}\ \emph {et~al.}(2022)\citenamefont {Wang},
  \citenamefont {Wang}, \citenamefont {Wang},\ and\ \citenamefont
  {Ren}}]{Wang_2022}%
  \BibitemOpen
  \bibfield  {author} {\bibinfo {author} {\bibfnamefont {L.}~\bibnamefont
  {Wang}}, \bibinfo {author} {\bibfnamefont {Z.}~\bibnamefont {Wang}}, \bibinfo
  {author} {\bibfnamefont {C.}~\bibnamefont {Wang}},\ and\ \bibinfo {author}
  {\bibfnamefont {J.}~\bibnamefont {Ren}},\ }\bibfield  {title} {\bibinfo
  {title} {{Cycle Flux Ranking of Network Analysis in Quantum Thermal
  Devices}},\ }\href {https://doi.org/10.1103/physrevlett.128.067701}
  {\bibfield  {journal} {\bibinfo  {journal} {Physical Review Letters}\
  }\textbf {\bibinfo {volume} {128}},\ \bibinfo {pages} {067701} (\bibinfo
  {year} {2022})}\BibitemShut {NoStop}%
\bibitem [{\citenamefont {Biehs}\ and\ \citenamefont
  {Agarwal}(2023)}]{Biehs_2023}%
  \BibitemOpen
  \bibfield  {author} {\bibinfo {author} {\bibfnamefont {S.-A.}\ \bibnamefont
  {Biehs}}\ and\ \bibinfo {author} {\bibfnamefont {G.~S.}\ \bibnamefont
  {Agarwal}},\ }\bibfield  {title} {\bibinfo {title} {{Breakdown of Detailed
  Balance for Thermal Radiation by Synthetic Fields}},\ }\href
  {https://doi.org/10.1103/PhysRevLett.130.110401} {\bibfield  {journal}
  {\bibinfo  {journal} {Physical Review Letters}\ }\textbf {\bibinfo {volume}
  {130}},\ \bibinfo {pages} {110401} (\bibinfo {year} {2023})}\BibitemShut
  {NoStop}%
\bibitem [{\citenamefont {Wang}\ \emph {et~al.}(2023)\citenamefont {Wang},
  \citenamefont {Zeng}, \citenamefont {Zhu}, \citenamefont {Wang},\ and\
  \citenamefont {Li}}]{Wang2023}%
  \BibitemOpen
  \bibfield  {author} {\bibinfo {author} {\bibfnamefont {S.}~\bibnamefont
  {Wang}}, \bibinfo {author} {\bibfnamefont {C.}~\bibnamefont {Zeng}}, \bibinfo
  {author} {\bibfnamefont {G.}~\bibnamefont {Zhu}}, \bibinfo {author}
  {\bibfnamefont {H.}~\bibnamefont {Wang}},\ and\ \bibinfo {author}
  {\bibfnamefont {B.}~\bibnamefont {Li}},\ }\bibfield  {title} {\bibinfo
  {title} {Controlling heat ratchet and flow reversal with simple networks},\
  }\href {https://doi.org/10.1103/PhysRevResearch.5.043009} {\bibfield
  {journal} {\bibinfo  {journal} {Phys. Rev. Res.}\ }\textbf {\bibinfo {volume}
  {5}},\ \bibinfo {pages} {043009} (\bibinfo {year} {2023})}\BibitemShut
  {NoStop}%
\bibitem [{\citenamefont {Horn}(1972)}]{Horn_1972}%
  \BibitemOpen
  \bibfield  {author} {\bibinfo {author} {\bibfnamefont {F.}~\bibnamefont
  {Horn}},\ }\bibfield  {title} {\bibinfo {title} {Necessary and sufficient
  conditions for complex balancing in chemical kinetics},\ }\href
  {https://doi.org/10.1007/bf00255664} {\bibfield  {journal} {\bibinfo
  {journal} {Archive for Rational Mechanics and Analysis}\ }\textbf {\bibinfo
  {volume} {49}},\ \bibinfo {pages} {172} (\bibinfo {year} {1972})}\BibitemShut
  {NoStop}%
\bibitem [{\citenamefont {Feinberg}(1972)}]{Feinberg_1972}%
  \BibitemOpen
  \bibfield  {author} {\bibinfo {author} {\bibfnamefont {M.}~\bibnamefont
  {Feinberg}},\ }\bibfield  {title} {\bibinfo {title} {Complex balancing in
  general kinetic systems},\ }\href {https://doi.org/10.1007/bf00255665}
  {\bibfield  {journal} {\bibinfo  {journal} {Archive for Rational Mechanics
  and Analysis}\ }\textbf {\bibinfo {volume} {49}},\ \bibinfo {pages} {187}
  (\bibinfo {year} {1972})}\BibitemShut {NoStop}%
\bibitem [{\citenamefont {Caloz}\ \emph {et~al.}(2018)\citenamefont {Caloz},
  \citenamefont {Al\`u}, \citenamefont {Tretyakov}, \citenamefont {Sounas},
  \citenamefont {Achouri},\ and\ \citenamefont {Deck-L\'eger}}]{Caloz_2018}%
  \BibitemOpen
  \bibfield  {author} {\bibinfo {author} {\bibfnamefont {C.}~\bibnamefont
  {Caloz}}, \bibinfo {author} {\bibfnamefont {A.}~\bibnamefont {Al\`u}},
  \bibinfo {author} {\bibfnamefont {S.}~\bibnamefont {Tretyakov}}, \bibinfo
  {author} {\bibfnamefont {D.}~\bibnamefont {Sounas}}, \bibinfo {author}
  {\bibfnamefont {K.}~\bibnamefont {Achouri}},\ and\ \bibinfo {author}
  {\bibfnamefont {Z.-L.}\ \bibnamefont {Deck-L\'eger}},\ }\bibfield  {title}
  {\bibinfo {title} {{Electromagnetic Nonreciprocity}},\ }\href
  {https://doi.org/10.1103/PhysRevApplied.10.047001} {\bibfield  {journal}
  {\bibinfo  {journal} {Physical Review Applied}\ }\textbf {\bibinfo {volume}
  {10}},\ \bibinfo {pages} {047001} (\bibinfo {year} {2018})}\BibitemShut
  {NoStop}%
\bibitem [{\citenamefont {Gnesotto}\ \emph {et~al.}(2018)\citenamefont
  {Gnesotto}, \citenamefont {Mura}, \citenamefont {Gladrow},\ and\
  \citenamefont {Broedersz}}]{Gnesotto2018}%
  \BibitemOpen
  \bibfield  {author} {\bibinfo {author} {\bibfnamefont {F.~S.}\ \bibnamefont
  {Gnesotto}}, \bibinfo {author} {\bibfnamefont {F.}~\bibnamefont {Mura}},
  \bibinfo {author} {\bibfnamefont {J.}~\bibnamefont {Gladrow}},\ and\ \bibinfo
  {author} {\bibfnamefont {C.~P.}\ \bibnamefont {Broedersz}},\ }\bibfield
  {title} {\bibinfo {title} {{Broken Detailed Balance and Non-equilibrium
  Dynamics in Living Systems: a Review}},\ }\href
  {https://doi.org/10.1088/1361-6633/aab3ed} {\bibfield  {journal} {\bibinfo
  {journal} {Reports on Progress in Physics}\ }\textbf {\bibinfo {volume}
  {81}},\ \bibinfo {pages} {066601} (\bibinfo {year} {2018})}\BibitemShut
  {NoStop}%
\bibitem [{\citenamefont {Mart{\'\i}nez}\ \emph {et~al.}(2019)\citenamefont
  {Mart{\'\i}nez}, \citenamefont {Bisker}, \citenamefont {Horowitz},\ and\
  \citenamefont {Parrondo}}]{Martnez_2019}%
  \BibitemOpen
  \bibfield  {author} {\bibinfo {author} {\bibfnamefont {I.~A.}\ \bibnamefont
  {Mart{\'\i}nez}}, \bibinfo {author} {\bibfnamefont {G.}~\bibnamefont
  {Bisker}}, \bibinfo {author} {\bibfnamefont {J.~M.}\ \bibnamefont
  {Horowitz}},\ and\ \bibinfo {author} {\bibfnamefont {J.~M.~R.}\ \bibnamefont
  {Parrondo}},\ }\bibfield  {title} {\bibinfo {title} {{Inferring Broken
  Detailed Balance in the Absence of Observable Currents}},\ }\href
  {https://doi.org/10.1038/s41467-019-11051-w} {\bibfield  {journal} {\bibinfo
  {journal} {Nature Communications}\ }\textbf {\bibinfo {volume} {10}},\
  \bibinfo {pages} {3542} (\bibinfo {year} {2019})}\BibitemShut {NoStop}%
\bibitem [{\citenamefont {Yu}\ \emph {et~al.}(2024)\citenamefont {Yu},
  \citenamefont {Zou}, \citenamefont {Zeng}, \citenamefont {Rao},\ and\
  \citenamefont {Xia}}]{Yu_2024}%
  \BibitemOpen
  \bibfield  {author} {\bibinfo {author} {\bibfnamefont {T.}~\bibnamefont
  {Yu}}, \bibinfo {author} {\bibfnamefont {J.}~\bibnamefont {Zou}}, \bibinfo
  {author} {\bibfnamefont {B.}~\bibnamefont {Zeng}}, \bibinfo {author}
  {\bibfnamefont {J.}~\bibnamefont {Rao}},\ and\ \bibinfo {author}
  {\bibfnamefont {K.}~\bibnamefont {Xia}},\ }\bibfield  {title} {\bibinfo
  {title} {{Non-Hermitian Topological Magnonics}},\ }\href
  {https://doi.org/https://doi.org/10.1016/j.physrep.2024.01.006} {\bibfield
  {journal} {\bibinfo  {journal} {Physics Reports}\ }\textbf {\bibinfo {volume}
  {1062}},\ \bibinfo {pages} {1} (\bibinfo {year} {2024})}\BibitemShut
  {NoStop}%
\bibitem [{\citenamefont {Shi}\ \emph {et~al.}(2015)\citenamefont {Shi},
  \citenamefont {Chang},\ and\ \citenamefont {Cirac}}]{Shi2015}%
  \BibitemOpen
  \bibfield  {author} {\bibinfo {author} {\bibfnamefont {T.}~\bibnamefont
  {Shi}}, \bibinfo {author} {\bibfnamefont {D.~E.}\ \bibnamefont {Chang}},\
  and\ \bibinfo {author} {\bibfnamefont {J.~I.}\ \bibnamefont {Cirac}},\
  }\bibfield  {title} {\bibinfo {title} {Multiphoton-scattering theory and
  generalized master equations},\ }\href
  {https://doi.org/10.1103/PhysRevA.92.053834} {\bibfield  {journal} {\bibinfo
  {journal} {Phys. Rev. A}\ }\textbf {\bibinfo {volume} {92}},\ \bibinfo
  {pages} {053834} (\bibinfo {year} {2015})}\BibitemShut {NoStop}%
\bibitem [{\citenamefont {Manzano}\ \emph {et~al.}(2013)\citenamefont
  {Manzano}, \citenamefont {Galve}, \citenamefont {Giorgi}, \citenamefont
  {Hern{\'a}ndez-Garc{\'\i}a},\ and\ \citenamefont {Zambrini}}]{Manzano2013}%
  \BibitemOpen
  \bibfield  {author} {\bibinfo {author} {\bibfnamefont {G.}~\bibnamefont
  {Manzano}}, \bibinfo {author} {\bibfnamefont {F.}~\bibnamefont {Galve}},
  \bibinfo {author} {\bibfnamefont {G.~L.}\ \bibnamefont {Giorgi}}, \bibinfo
  {author} {\bibfnamefont {E.}~\bibnamefont {Hern{\'a}ndez-Garc{\'\i}a}},\ and\
  \bibinfo {author} {\bibfnamefont {R.}~\bibnamefont {Zambrini}},\ }\bibfield
  {title} {\bibinfo {title} {Synchronization, quantum correlations and
  entanglement in oscillator networks},\ }\href@noop {} {\bibfield  {journal}
  {\bibinfo  {journal} {Scientific Reports}\ }\textbf {\bibinfo {volume} {3}},\
  \bibinfo {pages} {1439} (\bibinfo {year} {2013})}\BibitemShut {NoStop}%
\bibitem [{\citenamefont {Koppenh\"ofer}\ \emph {et~al.}(2020)\citenamefont
  {Koppenh\"ofer}, \citenamefont {Bruder},\ and\ \citenamefont
  {Roulet}}]{Koppenhofer2020}%
  \BibitemOpen
  \bibfield  {author} {\bibinfo {author} {\bibfnamefont {M.}~\bibnamefont
  {Koppenh\"ofer}}, \bibinfo {author} {\bibfnamefont {C.}~\bibnamefont
  {Bruder}},\ and\ \bibinfo {author} {\bibfnamefont {A.}~\bibnamefont
  {Roulet}},\ }\bibfield  {title} {\bibinfo {title} {Quantum synchronization on
  the ibm q system},\ }\href {https://doi.org/10.1103/PhysRevResearch.2.023026}
  {\bibfield  {journal} {\bibinfo  {journal} {Phys. Rev. Res.}\ }\textbf
  {\bibinfo {volume} {2}},\ \bibinfo {pages} {023026} (\bibinfo {year}
  {2020})}\BibitemShut {NoStop}%
\bibitem [{\citenamefont {Schmolke}\ and\ \citenamefont
  {Lutz}(2022)}]{Schmolke2022}%
  \BibitemOpen
  \bibfield  {author} {\bibinfo {author} {\bibfnamefont {F.}~\bibnamefont
  {Schmolke}}\ and\ \bibinfo {author} {\bibfnamefont {E.}~\bibnamefont
  {Lutz}},\ }\bibfield  {title} {\bibinfo {title} {Noise-induced quantum
  synchronization},\ }\href {https://doi.org/10.1103/PhysRevLett.129.250601}
  {\bibfield  {journal} {\bibinfo  {journal} {Phys. Rev. Lett.}\ }\textbf
  {\bibinfo {volume} {129}},\ \bibinfo {pages} {250601} (\bibinfo {year}
  {2022})}\BibitemShut {NoStop}%
\bibitem [{\citenamefont {Lynn}\ \emph {et~al.}(2021)\citenamefont {Lynn},
  \citenamefont {Cornblath}, \citenamefont {Papadopoulos}, \citenamefont
  {Bertolero},\ and\ \citenamefont {Bassett}}]{Lynn2021}%
  \BibitemOpen
  \bibfield  {author} {\bibinfo {author} {\bibfnamefont {C.~W.}\ \bibnamefont
  {Lynn}}, \bibinfo {author} {\bibfnamefont {E.~J.}\ \bibnamefont {Cornblath}},
  \bibinfo {author} {\bibfnamefont {L.}~\bibnamefont {Papadopoulos}}, \bibinfo
  {author} {\bibfnamefont {M.~A.}\ \bibnamefont {Bertolero}},\ and\ \bibinfo
  {author} {\bibfnamefont {D.~S.}\ \bibnamefont {Bassett}},\ }\bibfield
  {title} {\bibinfo {title} {{Broken Detailed Balance and Entropy Production in
  the Human Brain}},\ }\href {https://doi.org/10.1073/pnas.2109889118}
  {\bibfield  {journal} {\bibinfo  {journal} {Proceedings of the National
  Academy of Sciences}\ }\textbf {\bibinfo {volume} {118}},\ \bibinfo {pages}
  {066601} (\bibinfo {year} {2021})}\BibitemShut {NoStop}%
\bibitem [{\citenamefont {Nartallo-Kaluarachchi}\ \emph
  {et~al.}(2024)\citenamefont {Nartallo-Kaluarachchi}, \citenamefont {Asllani},
  \citenamefont {Deco}, \citenamefont {Kringelbach}, \citenamefont {Goriely},\
  and\ \citenamefont {Lambiotte}}]{Nartallo_Kaluarachchi_2024}%
  \BibitemOpen
  \bibfield  {author} {\bibinfo {author} {\bibfnamefont {R.}~\bibnamefont
  {Nartallo-Kaluarachchi}}, \bibinfo {author} {\bibfnamefont {M.}~\bibnamefont
  {Asllani}}, \bibinfo {author} {\bibfnamefont {G.}~\bibnamefont {Deco}},
  \bibinfo {author} {\bibfnamefont {M.~L.}\ \bibnamefont {Kringelbach}},
  \bibinfo {author} {\bibfnamefont {A.}~\bibnamefont {Goriely}},\ and\ \bibinfo
  {author} {\bibfnamefont {R.}~\bibnamefont {Lambiotte}},\ }\bibfield  {title}
  {\bibinfo {title} {{Broken Detailed Balance and Entropy Production in
  Directed Networks}},\ }\href {https://doi.org/10.1103/physreve.110.034313}
  {\bibfield  {journal} {\bibinfo  {journal} {Physical Review E}\ }\textbf
  {\bibinfo {volume} {110}},\ \bibinfo {pages} {034313} (\bibinfo {year}
  {2024})}\BibitemShut {NoStop}%
\bibitem [{\citenamefont {Monti}\ \emph {et~al.}(2025)\citenamefont {Monti},
  \citenamefont {Perl}, \citenamefont {Tagliazucchi}, \citenamefont
  {Kringelbach},\ and\ \citenamefont {Deco}}]{Monti2025}%
  \BibitemOpen
  \bibfield  {author} {\bibinfo {author} {\bibfnamefont {J.~M.}\ \bibnamefont
  {Monti}}, \bibinfo {author} {\bibfnamefont {Y.~S.}\ \bibnamefont {Perl}},
  \bibinfo {author} {\bibfnamefont {E.}~\bibnamefont {Tagliazucchi}}, \bibinfo
  {author} {\bibfnamefont {M.~L.}\ \bibnamefont {Kringelbach}},\ and\ \bibinfo
  {author} {\bibfnamefont {G.}~\bibnamefont {Deco}},\ }\bibfield  {title}
  {\bibinfo {title} {{Fluctuation-dissipation theorem and the discovery of
  distinctive off-equilibrium signatures of brain states}},\ }\href
  {https://doi.org/10.1103/PhysRevResearch.7.013301} {\bibfield  {journal}
  {\bibinfo  {journal} {Physical Review Research}\ }\textbf {\bibinfo {volume}
  {7}},\ \bibinfo {pages} {013301} (\bibinfo {year} {2025})}\BibitemShut
  {NoStop}%
\bibitem [{\citenamefont {Gladrow}\ \emph {et~al.}(2016)\citenamefont
  {Gladrow}, \citenamefont {Fakhri}, \citenamefont {MacKintosh}, \citenamefont
  {Schmidt},\ and\ \citenamefont {Broedersz}}]{Gladrow2016}%
  \BibitemOpen
  \bibfield  {author} {\bibinfo {author} {\bibfnamefont {J.}~\bibnamefont
  {Gladrow}}, \bibinfo {author} {\bibfnamefont {N.}~\bibnamefont {Fakhri}},
  \bibinfo {author} {\bibfnamefont {F.~C.}\ \bibnamefont {MacKintosh}},
  \bibinfo {author} {\bibfnamefont {C.~F.}\ \bibnamefont {Schmidt}},\ and\
  \bibinfo {author} {\bibfnamefont {C.~P.}\ \bibnamefont {Broedersz}},\
  }\bibfield  {title} {\bibinfo {title} {{Broken Detailed Balance of Filament
  Dynamics in Active Networks}},\ }\href
  {https://doi.org/10.1103/PhysRevLett.116.248301} {\bibfield  {journal}
  {\bibinfo  {journal} {Physical Review Letters}\ }\textbf {\bibinfo {volume}
  {116}},\ \bibinfo {pages} {248301} (\bibinfo {year} {2016})}\BibitemShut
  {NoStop}%
\bibitem [{\citenamefont {Battle}\ \emph {et~al.}(2016)\citenamefont {Battle},
  \citenamefont {Broedersz}, \citenamefont {Fakhri}, \citenamefont {Geyer},
  \citenamefont {Howard}, \citenamefont {Schmidt},\ and\ \citenamefont
  {MacKintosh}}]{Battle2016}%
  \BibitemOpen
  \bibfield  {author} {\bibinfo {author} {\bibfnamefont {C.}~\bibnamefont
  {Battle}}, \bibinfo {author} {\bibfnamefont {C.~P.}\ \bibnamefont
  {Broedersz}}, \bibinfo {author} {\bibfnamefont {N.}~\bibnamefont {Fakhri}},
  \bibinfo {author} {\bibfnamefont {V.~F.}\ \bibnamefont {Geyer}}, \bibinfo
  {author} {\bibfnamefont {J.}~\bibnamefont {Howard}}, \bibinfo {author}
  {\bibfnamefont {C.~F.}\ \bibnamefont {Schmidt}},\ and\ \bibinfo {author}
  {\bibfnamefont {F.~C.}\ \bibnamefont {MacKintosh}},\ }\bibfield  {title}
  {\bibinfo {title} {{Broken Detailed Balance at Mesoscopic Scales in Active
  Biological Systems}},\ }\href {https://doi.org/10.1126/science.aac8167}
  {\bibfield  {journal} {\bibinfo  {journal} {Science}\ }\textbf {\bibinfo
  {volume} {352}},\ \bibinfo {pages} {604} (\bibinfo {year}
  {2016})}\BibitemShut {NoStop}%
\bibitem [{\citenamefont {Denisov}\ \emph {et~al.}(2002)\citenamefont
  {Denisov}, \citenamefont {Castro-Beltran},\ and\ \citenamefont
  {Carmichael}}]{Denisov_2002}%
  \BibitemOpen
  \bibfield  {author} {\bibinfo {author} {\bibfnamefont {A.}~\bibnamefont
  {Denisov}}, \bibinfo {author} {\bibfnamefont {H.~M.}\ \bibnamefont
  {Castro-Beltran}},\ and\ \bibinfo {author} {\bibfnamefont {H.~J.}\
  \bibnamefont {Carmichael}},\ }\bibfield  {title} {\bibinfo {title}
  {{Time-Asymmetric Fluctuations of Light and the Breakdown of Detailed
  Balance}},\ }\href {https://doi.org/10.1103/PhysRevLett.88.243601} {\bibfield
   {journal} {\bibinfo  {journal} {Physical Review Letters}\ }\textbf {\bibinfo
  {volume} {88}},\ \bibinfo {pages} {243601} (\bibinfo {year}
  {2002})}\BibitemShut {NoStop}%
\bibitem [{\citenamefont {S\'anchez}\ \emph {et~al.}(2010)\citenamefont
  {S\'anchez}, \citenamefont {L\'opez}, \citenamefont {S\'anchez},\ and\
  \citenamefont {B\"uttiker}}]{Sanchez_2010}%
  \BibitemOpen
  \bibfield  {author} {\bibinfo {author} {\bibfnamefont {R.}~\bibnamefont
  {S\'anchez}}, \bibinfo {author} {\bibfnamefont {R.}~\bibnamefont {L\'opez}},
  \bibinfo {author} {\bibfnamefont {D.}~\bibnamefont {S\'anchez}},\ and\
  \bibinfo {author} {\bibfnamefont {M.}~\bibnamefont {B\"uttiker}},\ }\bibfield
   {title} {\bibinfo {title} {{Mesoscopic Coulomb Drag, Broken Detailed
  Balance, and Fluctuation Relations}},\ }\href
  {https://doi.org/10.1103/PhysRevLett.104.076801} {\bibfield  {journal}
  {\bibinfo  {journal} {Physical Review Letters}\ }\textbf {\bibinfo {volume}
  {104}},\ \bibinfo {pages} {076801} (\bibinfo {year} {2010})}\BibitemShut
  {NoStop}%
\bibitem [{\citenamefont {Kim}\ \emph {et~al.}(2018)\citenamefont {Kim},
  \citenamefont {Park}, \citenamefont {Kim}, \citenamefont {Sim},\ and\
  \citenamefont {Ahn}}]{Kim_2018}%
  \BibitemOpen
  \bibfield  {author} {\bibinfo {author} {\bibfnamefont {H.}~\bibnamefont
  {Kim}}, \bibinfo {author} {\bibfnamefont {Y.}~\bibnamefont {Park}}, \bibinfo
  {author} {\bibfnamefont {K.}~\bibnamefont {Kim}}, \bibinfo {author}
  {\bibfnamefont {H.-S.}\ \bibnamefont {Sim}},\ and\ \bibinfo {author}
  {\bibfnamefont {J.}~\bibnamefont {Ahn}},\ }\bibfield  {title} {\bibinfo
  {title} {{Detailed Balance of Thermalization Dynamics in Rydberg-Atom Quantum
  Simulators}},\ }\href {https://doi.org/10.1103/PhysRevLett.120.180502}
  {\bibfield  {journal} {\bibinfo  {journal} {Physical Review Letters}\
  }\textbf {\bibinfo {volume} {120}},\ \bibinfo {pages} {180502} (\bibinfo
  {year} {2018})}\BibitemShut {NoStop}%
\bibitem [{\citenamefont {Breuer}\ and\ \citenamefont
  {Petruccione}(2007)}]{Breuer_2007}%
  \BibitemOpen
  \bibfield  {author} {\bibinfo {author} {\bibfnamefont {H.-P.}\ \bibnamefont
  {Breuer}}\ and\ \bibinfo {author} {\bibfnamefont {F.}~\bibnamefont
  {Petruccione}},\ }\href
  {https://doi.org/10.1093/acprof:oso/9780199213900.001.0001} {\emph {\bibinfo
  {title} {{The Theory of Open Quantum Systems}}}}\ (\bibinfo  {publisher}
  {Oxford University PressOxford},\ \bibinfo {year} {2007})\BibitemShut
  {NoStop}%
\bibitem [{\citenamefont {Weiss}(2011)}]{Weiss_2011}%
  \BibitemOpen
  \bibfield  {author} {\bibinfo {author} {\bibfnamefont {U.}~\bibnamefont
  {Weiss}},\ }\href {https://doi.org/10.1142/8334} {\emph {\bibinfo {title}
  {{Quantum Dissipative Systems}}}}\ (\bibinfo  {publisher} {WORLD
  SCIENTIFIC},\ \bibinfo {year} {2011})\BibitemShut {NoStop}%
\bibitem [{\citenamefont {Ichiki}\ and\ \citenamefont
  {Ohzeki}(2013)}]{Ichiki_2013}%
  \BibitemOpen
  \bibfield  {author} {\bibinfo {author} {\bibfnamefont {A.}~\bibnamefont
  {Ichiki}}\ and\ \bibinfo {author} {\bibfnamefont {M.}~\bibnamefont
  {Ohzeki}},\ }\bibfield  {title} {\bibinfo {title} {{Violation of Detailed
  Balance Accelerates Relaxation}},\ }\href
  {https://doi.org/10.1103/PhysRevE.88.020101} {\bibfield  {journal} {\bibinfo
  {journal} {Physical Review E}\ }\textbf {\bibinfo {volume} {88}},\ \bibinfo
  {pages} {020101} (\bibinfo {year} {2013})}\BibitemShut {NoStop}%
\bibitem [{\citenamefont {Breuer}\ \emph {et~al.}(2016)\citenamefont {Breuer},
  \citenamefont {Laine}, \citenamefont {Piilo},\ and\ \citenamefont
  {Vacchini}}]{Breuer_2016}%
  \BibitemOpen
  \bibfield  {author} {\bibinfo {author} {\bibfnamefont {H.-P.}\ \bibnamefont
  {Breuer}}, \bibinfo {author} {\bibfnamefont {E.-M.}\ \bibnamefont {Laine}},
  \bibinfo {author} {\bibfnamefont {J.}~\bibnamefont {Piilo}},\ and\ \bibinfo
  {author} {\bibfnamefont {B.}~\bibnamefont {Vacchini}},\ }\bibfield  {title}
  {\bibinfo {title} {Colloquium: Non-markovian dynamics in open quantum
  systems},\ }\href {https://doi.org/10.1103/revmodphys.88.021002} {\bibfield
  {journal} {\bibinfo  {journal} {Reviews of Modern Physics}\ }\textbf
  {\bibinfo {volume} {88}},\ \bibinfo {pages} {021002} (\bibinfo {year}
  {2016})}\BibitemShut {NoStop}%
\bibitem [{\citenamefont {Kjaergaard}\ \emph {et~al.}(2020)\citenamefont
  {Kjaergaard}, \citenamefont {Schwartz}, \citenamefont {Braum\"{u}ller},
  \citenamefont {Krantz}, \citenamefont {Wang}, \citenamefont {Gustavsson},\
  and\ \citenamefont {Oliver}}]{Kjaergaard2020}%
  \BibitemOpen
  \bibfield  {author} {\bibinfo {author} {\bibfnamefont {M.}~\bibnamefont
  {Kjaergaard}}, \bibinfo {author} {\bibfnamefont {M.~E.}\ \bibnamefont
  {Schwartz}}, \bibinfo {author} {\bibfnamefont {J.}~\bibnamefont
  {Braum\"{u}ller}}, \bibinfo {author} {\bibfnamefont {P.}~\bibnamefont
  {Krantz}}, \bibinfo {author} {\bibfnamefont {J.~I.-J.}\ \bibnamefont {Wang}},
  \bibinfo {author} {\bibfnamefont {S.}~\bibnamefont {Gustavsson}},\ and\
  \bibinfo {author} {\bibfnamefont {W.~D.}\ \bibnamefont {Oliver}},\ }\bibfield
   {title} {\bibinfo {title} {Superconducting qubits: Current state of play},\
  }\href {https://doi.org/10.1146/annurev-conmatphys-031119-050605} {\bibfield
  {journal} {\bibinfo  {journal} {Annual Review of Condensed Matter Physics}\
  }\textbf {\bibinfo {volume} {11}},\ \bibinfo {pages} {369} (\bibinfo {year}
  {2020})}\BibitemShut {NoStop}%
\bibitem [{\citenamefont {Krantz}\ \emph {et~al.}(2019)\citenamefont {Krantz},
  \citenamefont {Kjaergaard}, \citenamefont {Yan}, \citenamefont {Orlando},
  \citenamefont {Gustavsson},\ and\ \citenamefont {Oliver}}]{Krantz2019}%
  \BibitemOpen
  \bibfield  {author} {\bibinfo {author} {\bibfnamefont {P.}~\bibnamefont
  {Krantz}}, \bibinfo {author} {\bibfnamefont {M.}~\bibnamefont {Kjaergaard}},
  \bibinfo {author} {\bibfnamefont {F.}~\bibnamefont {Yan}}, \bibinfo {author}
  {\bibfnamefont {T.~P.}\ \bibnamefont {Orlando}}, \bibinfo {author}
  {\bibfnamefont {S.}~\bibnamefont {Gustavsson}},\ and\ \bibinfo {author}
  {\bibfnamefont {W.~D.}\ \bibnamefont {Oliver}},\ }\bibfield  {title}
  {\bibinfo {title} {A quantum engineer's guide to superconducting qubits},\
  }\bibfield  {journal} {\bibinfo  {journal} {Applied Physics Reviews}\
  }\textbf {\bibinfo {volume} {6}},\ \href {https://doi.org/10.1063/1.5089550}
  {10.1063/1.5089550} (\bibinfo {year} {2019})\BibitemShut {NoStop}%
\bibitem [{\citenamefont {Shi}\ \emph {et~al.}(2023)\citenamefont {Shi},
  \citenamefont {Liu}, \citenamefont {Zhang}, \citenamefont {Xiang},
  \citenamefont {Huang}, \citenamefont {Liu}, \citenamefont {Wang},
  \citenamefont {Zhang}, \citenamefont {Deng}, \citenamefont {Liang},
  \citenamefont {Mei}, \citenamefont {Li}, \citenamefont {Li}, \citenamefont
  {Ma}, \citenamefont {Liu}, \citenamefont {Chen}, \citenamefont {Liu},
  \citenamefont {Tian}, \citenamefont {Song}, \citenamefont {Zhao},
  \citenamefont {Xu}, \citenamefont {Zheng}, \citenamefont {Nori},\ and\
  \citenamefont {Fan}}]{Shi2023}%
  \BibitemOpen
  \bibfield  {author} {\bibinfo {author} {\bibfnamefont {Y.-H.}\ \bibnamefont
  {Shi}}, \bibinfo {author} {\bibfnamefont {Y.}~\bibnamefont {Liu}}, \bibinfo
  {author} {\bibfnamefont {Y.-R.}\ \bibnamefont {Zhang}}, \bibinfo {author}
  {\bibfnamefont {Z.}~\bibnamefont {Xiang}}, \bibinfo {author} {\bibfnamefont
  {K.}~\bibnamefont {Huang}}, \bibinfo {author} {\bibfnamefont
  {T.}~\bibnamefont {Liu}}, \bibinfo {author} {\bibfnamefont {Y.-Y.}\
  \bibnamefont {Wang}}, \bibinfo {author} {\bibfnamefont {J.-C.}\ \bibnamefont
  {Zhang}}, \bibinfo {author} {\bibfnamefont {C.-L.}\ \bibnamefont {Deng}},
  \bibinfo {author} {\bibfnamefont {G.-H.}\ \bibnamefont {Liang}}, \bibinfo
  {author} {\bibfnamefont {Z.-Y.}\ \bibnamefont {Mei}}, \bibinfo {author}
  {\bibfnamefont {H.}~\bibnamefont {Li}}, \bibinfo {author} {\bibfnamefont
  {T.-M.}\ \bibnamefont {Li}}, \bibinfo {author} {\bibfnamefont {W.-G.}\
  \bibnamefont {Ma}}, \bibinfo {author} {\bibfnamefont {H.-T.}\ \bibnamefont
  {Liu}}, \bibinfo {author} {\bibfnamefont {C.-T.}\ \bibnamefont {Chen}},
  \bibinfo {author} {\bibfnamefont {T.}~\bibnamefont {Liu}}, \bibinfo {author}
  {\bibfnamefont {Y.}~\bibnamefont {Tian}}, \bibinfo {author} {\bibfnamefont
  {X.}~\bibnamefont {Song}}, \bibinfo {author} {\bibfnamefont {S.~P.}\
  \bibnamefont {Zhao}}, \bibinfo {author} {\bibfnamefont {K.}~\bibnamefont
  {Xu}}, \bibinfo {author} {\bibfnamefont {D.}~\bibnamefont {Zheng}}, \bibinfo
  {author} {\bibfnamefont {F.}~\bibnamefont {Nori}},\ and\ \bibinfo {author}
  {\bibfnamefont {H.}~\bibnamefont {Fan}},\ }\bibfield  {title} {\bibinfo
  {title} {Quantum simulation of topological zero modes on a 41-qubit
  superconducting processor},\ }\href
  {https://doi.org/10.1103/PhysRevLett.131.080401} {\bibfield  {journal}
  {\bibinfo  {journal} {Phys. Rev. Lett.}\ }\textbf {\bibinfo {volume} {131}},\
  \bibinfo {pages} {080401} (\bibinfo {year} {2023})}\BibitemShut {NoStop}%
\bibitem [{\citenamefont {Manovitz}\ \emph {et~al.}(2025)\citenamefont
  {Manovitz}, \citenamefont {Li}, \citenamefont {Ebadi}, \citenamefont
  {Samajdar}, \citenamefont {Geim}, \citenamefont {Evered}, \citenamefont
  {Bluvstein}, \citenamefont {Zhou}, \citenamefont {Koyluoglu}, \citenamefont
  {Feldmeier} \emph {et~al.}}]{Manovitz2025}%
  \BibitemOpen
  \bibfield  {author} {\bibinfo {author} {\bibfnamefont {T.}~\bibnamefont
  {Manovitz}}, \bibinfo {author} {\bibfnamefont {S.~H.}\ \bibnamefont {Li}},
  \bibinfo {author} {\bibfnamefont {S.}~\bibnamefont {Ebadi}}, \bibinfo
  {author} {\bibfnamefont {R.}~\bibnamefont {Samajdar}}, \bibinfo {author}
  {\bibfnamefont {A.~A.}\ \bibnamefont {Geim}}, \bibinfo {author}
  {\bibfnamefont {S.~J.}\ \bibnamefont {Evered}}, \bibinfo {author}
  {\bibfnamefont {D.}~\bibnamefont {Bluvstein}}, \bibinfo {author}
  {\bibfnamefont {H.}~\bibnamefont {Zhou}}, \bibinfo {author} {\bibfnamefont
  {N.~U.}\ \bibnamefont {Koyluoglu}}, \bibinfo {author} {\bibfnamefont
  {J.}~\bibnamefont {Feldmeier}}, \emph {et~al.},\ }\bibfield  {title}
  {\bibinfo {title} {Quantum coarsening and collective dynamics on a
  programmable simulator},\ }\href@noop {} {\bibfield  {journal} {\bibinfo
  {journal} {Nature}\ }\textbf {\bibinfo {volume} {638}},\ \bibinfo {pages}
  {86} (\bibinfo {year} {2025})}\BibitemShut {NoStop}%
\bibitem [{\citenamefont {Jing}\ and\ \citenamefont {Wu}(2018)}]{Jing_2018}%
  \BibitemOpen
  \bibfield  {author} {\bibinfo {author} {\bibfnamefont {J.}~\bibnamefont
  {Jing}}\ and\ \bibinfo {author} {\bibfnamefont {L.-A.}\ \bibnamefont {Wu}},\
  }\bibfield  {title} {\bibinfo {title} {Decoherence and control of a qubit in
  spin baths: an exact master equation study},\ }\href
  {https://doi.org/10.1038/s41598-018-19977-9} {\bibfield  {journal} {\bibinfo
  {journal} {Scientific Reports}\ }\textbf {\bibinfo {volume} {8}},\ \bibinfo
  {pages} {1471} (\bibinfo {year} {2018})}\BibitemShut {NoStop}%
\bibitem [{\citenamefont {Cusumano}(2022)}]{Cusumano_2022}%
  \BibitemOpen
  \bibfield  {author} {\bibinfo {author} {\bibfnamefont {S.}~\bibnamefont
  {Cusumano}},\ }\bibfield  {title} {\bibinfo {title} {Quantum collision
  models: A beginner guide},\ }\href {https://doi.org/10.3390/e24091258}
  {\bibfield  {journal} {\bibinfo  {journal} {Entropy}\ }\textbf {\bibinfo
  {volume} {24}},\ \bibinfo {pages} {1258} (\bibinfo {year}
  {2022})}\BibitemShut {NoStop}%
\bibitem [{\citenamefont {Di~Bartolomeo}\ \emph {et~al.}(2024)\citenamefont
  {Di~Bartolomeo}, \citenamefont {Vischi}, \citenamefont {Feri}, \citenamefont
  {Bassi},\ and\ \citenamefont {Donadi}}]{Di_Bartolomeo_2024}%
  \BibitemOpen
  \bibfield  {author} {\bibinfo {author} {\bibfnamefont {G.}~\bibnamefont
  {Di~Bartolomeo}}, \bibinfo {author} {\bibfnamefont {M.}~\bibnamefont
  {Vischi}}, \bibinfo {author} {\bibfnamefont {T.}~\bibnamefont {Feri}},
  \bibinfo {author} {\bibfnamefont {A.}~\bibnamefont {Bassi}},\ and\ \bibinfo
  {author} {\bibfnamefont {S.}~\bibnamefont {Donadi}},\ }\bibfield  {title}
  {\bibinfo {title} {Efficient quantum algorithm to simulate open systems
  through a single environmental qubit},\ }\href
  {https://doi.org/10.1103/physrevresearch.6.043321} {\bibfield  {journal}
  {\bibinfo  {journal} {Physical Review Research}\ }\textbf {\bibinfo {volume}
  {6}},\ \bibinfo {pages} {043321} (\bibinfo {year} {2024})}\BibitemShut
  {NoStop}%
\bibitem [{\citenamefont {G{\"u}nther}\ and\ \citenamefont
  {Samsonov}(2008)}]{G_nther_2008}%
  \BibitemOpen
  \bibfield  {author} {\bibinfo {author} {\bibfnamefont {U.}~\bibnamefont
  {G{\"u}nther}}\ and\ \bibinfo {author} {\bibfnamefont {B.~F.}\ \bibnamefont
  {Samsonov}},\ }\bibfield  {title} {\bibinfo {title} {Naimark-dilated
  $\mathcal{P}\mathcal{T}$-symmetric brachistochrone},\ }\href
  {https://doi.org/10.1103/PhysRevLett.101.230404} {\bibfield  {journal}
  {\bibinfo  {journal} {Physical Review Letters}\ }\textbf {\bibinfo {volume}
  {101}},\ \bibinfo {pages} {230404} (\bibinfo {year} {2008})}\BibitemShut
  {NoStop}%
\bibitem [{\citenamefont {Chen}\ \emph {et~al.}(2021)\citenamefont {Chen},
  \citenamefont {Abbasi}, \citenamefont {Joglekar},\ and\ \citenamefont
  {Murch}}]{Chen_2021}%
  \BibitemOpen
  \bibfield  {author} {\bibinfo {author} {\bibfnamefont {W.}~\bibnamefont
  {Chen}}, \bibinfo {author} {\bibfnamefont {M.}~\bibnamefont {Abbasi}},
  \bibinfo {author} {\bibfnamefont {Y.~N.}\ \bibnamefont {Joglekar}},\ and\
  \bibinfo {author} {\bibfnamefont {K.~W.}\ \bibnamefont {Murch}},\ }\bibfield
  {title} {\bibinfo {title} {{Quantum Jumps in the Non-Hermitian Dynamics of a
  Superconducting Qubit}},\ }\href
  {https://doi.org/10.1103/PhysRevLett.127.140504} {\bibfield  {journal}
  {\bibinfo  {journal} {Physical Review Letters}\ }\textbf {\bibinfo {volume}
  {127}},\ \bibinfo {pages} {140504} (\bibinfo {year} {2021})}\BibitemShut
  {NoStop}%
\bibitem [{\citenamefont {Zhang}\ \emph {et~al.}(2021)\citenamefont {Zhang},
  \citenamefont {Liu},\ and\ \citenamefont {Yung}}]{Zhang_2021}%
  \BibitemOpen
  \bibfield  {author} {\bibinfo {author} {\bibfnamefont {G.-L.}\ \bibnamefont
  {Zhang}}, \bibinfo {author} {\bibfnamefont {D.}~\bibnamefont {Liu}},\ and\
  \bibinfo {author} {\bibfnamefont {M.-H.}\ \bibnamefont {Yung}},\ }\bibfield
  {title} {\bibinfo {title} {Observation of exceptional point in a pt broken
  non-hermitian system simulated using a quantum circuit},\ }\href
  {https://doi.org/10.1038/s41598-021-93192-x} {\bibfield  {journal} {\bibinfo
  {journal} {Scientific Reports}\ }\textbf {\bibinfo {volume} {11}},\ \bibinfo
  {pages} {13795} (\bibinfo {year} {2021})}\BibitemShut {NoStop}%
\bibitem [{\citenamefont {Jebraeilli}\ and\ \citenamefont
  {Geller}(2025)}]{Jebraeilli_2025}%
  \BibitemOpen
  \bibfield  {author} {\bibinfo {author} {\bibfnamefont {A.}~\bibnamefont
  {Jebraeilli}}\ and\ \bibinfo {author} {\bibfnamefont {M.~R.}\ \bibnamefont
  {Geller}},\ }\bibfield  {title} {\bibinfo {title} {Quantum simulation of a
  qubit with a non-hermitian hamiltonian},\ }\href
  {https://doi.org/10.1103/physreva.111.032211} {\bibfield  {journal} {\bibinfo
   {journal} {Physical Review A}\ }\textbf {\bibinfo {volume} {111}},\ \bibinfo
  {pages} {032211} (\bibinfo {year} {2025})}\BibitemShut {NoStop}%
\bibitem [{\citenamefont {R\"{u}ter}\ \emph {et~al.}(2010)\citenamefont
  {R\"{u}ter}, \citenamefont {Makris}, \citenamefont {El-Ganainy},
  \citenamefont {Christodoulides}, \citenamefont {Segev},\ and\ \citenamefont
  {Kip}}]{Ruter_2010}%
  \BibitemOpen
  \bibfield  {author} {\bibinfo {author} {\bibfnamefont {C.~E.}\ \bibnamefont
  {R\"{u}ter}}, \bibinfo {author} {\bibfnamefont {K.~G.}\ \bibnamefont
  {Makris}}, \bibinfo {author} {\bibfnamefont {R.}~\bibnamefont {El-Ganainy}},
  \bibinfo {author} {\bibfnamefont {D.~N.}\ \bibnamefont {Christodoulides}},
  \bibinfo {author} {\bibfnamefont {M.}~\bibnamefont {Segev}},\ and\ \bibinfo
  {author} {\bibfnamefont {D.}~\bibnamefont {Kip}},\ }\bibfield  {title}
  {\bibinfo {title} {{Observation of Parity--Time Symmetry in Optics}},\ }\href
  {https://doi.org/10.1038/nphys1515} {\bibfield  {journal} {\bibinfo
  {journal} {Nature Physics}\ }\textbf {\bibinfo {volume} {6}},\ \bibinfo
  {pages} {192} (\bibinfo {year} {2010})}\BibitemShut {NoStop}%
\bibitem [{\citenamefont {Naghiloo}\ \emph {et~al.}(2019)\citenamefont
  {Naghiloo}, \citenamefont {Abbasi}, \citenamefont {Joglekar},\ and\
  \citenamefont {Murch}}]{Naghiloo_2019}%
  \BibitemOpen
  \bibfield  {author} {\bibinfo {author} {\bibfnamefont {M.}~\bibnamefont
  {Naghiloo}}, \bibinfo {author} {\bibfnamefont {M.}~\bibnamefont {Abbasi}},
  \bibinfo {author} {\bibfnamefont {Y.~N.}\ \bibnamefont {Joglekar}},\ and\
  \bibinfo {author} {\bibfnamefont {K.~W.}\ \bibnamefont {Murch}},\ }\bibfield
  {title} {\bibinfo {title} {{Quantum State Tomography across the Exceptional
  Point in a Single Dissipative Qubit}},\ }\href
  {https://doi.org/10.1038/s41567-019-0652-z} {\bibfield  {journal} {\bibinfo
  {journal} {Nature Physics}\ }\textbf {\bibinfo {volume} {15}},\ \bibinfo
  {pages} {1232} (\bibinfo {year} {2019})}\BibitemShut {NoStop}%
\bibitem [{\citenamefont {Du}\ \emph {et~al.}(2022)\citenamefont {Du},
  \citenamefont {Cao},\ and\ \citenamefont {Kou}}]{Du_2022}%
  \BibitemOpen
  \bibfield  {author} {\bibinfo {author} {\bibfnamefont {Q.}~\bibnamefont
  {Du}}, \bibinfo {author} {\bibfnamefont {K.}~\bibnamefont {Cao}},\ and\
  \bibinfo {author} {\bibfnamefont {S.-P.}\ \bibnamefont {Kou}},\ }\bibfield
  {title} {\bibinfo {title} {{Physics of $\mathcal{PT}$-symmetric Quantum
  Systems at Finite Temperatures}},\ }\href
  {https://doi.org/10.1103/physreva.106.032206} {\bibfield  {journal} {\bibinfo
   {journal} {Physical Review A}\ }\textbf {\bibinfo {volume} {106}},\ \bibinfo
  {pages} {032206} (\bibinfo {year} {2022})}\BibitemShut {NoStop}%
\bibitem [{\citenamefont {Singha~Roy}\ \emph {et~al.}(2025)\citenamefont
  {Singha~Roy}, \citenamefont {Bandyopadhyay}, \citenamefont {Costa~de
  Almeida},\ and\ \citenamefont {Hauke}}]{Roy_2023}%
  \BibitemOpen
  \bibfield  {author} {\bibinfo {author} {\bibfnamefont {S.}~\bibnamefont
  {Singha~Roy}}, \bibinfo {author} {\bibfnamefont {S.}~\bibnamefont
  {Bandyopadhyay}}, \bibinfo {author} {\bibfnamefont {R.}~\bibnamefont
  {Costa~de Almeida}},\ and\ \bibinfo {author} {\bibfnamefont {P.}~\bibnamefont
  {Hauke}},\ }\bibfield  {title} {\bibinfo {title} {Unveiling eigenstate
  thermalization for non-hermitian systems},\ }\href
  {https://doi.org/10.1103/PhysRevLett.134.180405} {\bibfield  {journal}
  {\bibinfo  {journal} {Phys. Rev. Lett.}\ }\textbf {\bibinfo {volume} {134}},\
  \bibinfo {pages} {180405} (\bibinfo {year} {2025})}\BibitemShut {NoStop}%
\bibitem [{\citenamefont {Cipolloni}\ and\ \citenamefont
  {Kudler-Flam}(2024)}]{Cipolloni_2024}%
  \BibitemOpen
  \bibfield  {author} {\bibinfo {author} {\bibfnamefont {G.}~\bibnamefont
  {Cipolloni}}\ and\ \bibinfo {author} {\bibfnamefont {J.}~\bibnamefont
  {Kudler-Flam}},\ }\bibfield  {title} {\bibinfo {title} {{Non-Hermitian
  Hamiltonians Violate the Eigenstate Thermalization Hypothesis}},\ }\href
  {https://doi.org/10.1103/physrevb.109.l020201} {\bibfield  {journal}
  {\bibinfo  {journal} {Physical Review B}\ }\textbf {\bibinfo {volume}
  {109}},\ \bibinfo {pages} {L020201} (\bibinfo {year} {2024})}\BibitemShut
  {NoStop}%
\bibitem [{\citenamefont {Mao}\ \emph {et~al.}(2024)\citenamefont {Mao},
  \citenamefont {Zhong}, \citenamefont {Lin}, \citenamefont {Wang},\ and\
  \citenamefont {Hu}}]{Mao_2024}%
  \BibitemOpen
  \bibfield  {author} {\bibinfo {author} {\bibfnamefont {Y.}~\bibnamefont
  {Mao}}, \bibinfo {author} {\bibfnamefont {P.}~\bibnamefont {Zhong}}, \bibinfo
  {author} {\bibfnamefont {H.}~\bibnamefont {Lin}}, \bibinfo {author}
  {\bibfnamefont {X.}~\bibnamefont {Wang}},\ and\ \bibinfo {author}
  {\bibfnamefont {S.}~\bibnamefont {Hu}},\ }\bibfield  {title} {\bibinfo
  {title} {{Diagnosing Thermalization Dynamics of Non-Hermitian Quantum Systems
  via GKSL Master Equations}},\ }\href
  {https://doi.org/10.1088/0256-307x/41/7/070301} {\bibfield  {journal}
  {\bibinfo  {journal} {Chinese Physics Letters}\ }\textbf {\bibinfo {volume}
  {41}},\ \bibinfo {pages} {070301} (\bibinfo {year} {2024})}\BibitemShut
  {NoStop}%
\bibitem [{\citenamefont {Blok}\ \emph {et~al.}(2021)\citenamefont {Blok},
  \citenamefont {Ramasesh}, \citenamefont {Schuster}, \citenamefont {O'Brien},
  \citenamefont {Kreikebaum}, \citenamefont {Dahlen}, \citenamefont {Morvan},
  \citenamefont {Yoshida}, \citenamefont {Yao},\ and\ \citenamefont
  {Siddiqi}}]{Blok_2021}%
  \BibitemOpen
  \bibfield  {author} {\bibinfo {author} {\bibfnamefont {M.}~\bibnamefont
  {Blok}}, \bibinfo {author} {\bibfnamefont {V.}~\bibnamefont {Ramasesh}},
  \bibinfo {author} {\bibfnamefont {T.}~\bibnamefont {Schuster}}, \bibinfo
  {author} {\bibfnamefont {K.}~\bibnamefont {O'Brien}}, \bibinfo {author}
  {\bibfnamefont {J.}~\bibnamefont {Kreikebaum}}, \bibinfo {author}
  {\bibfnamefont {D.}~\bibnamefont {Dahlen}}, \bibinfo {author} {\bibfnamefont
  {A.}~\bibnamefont {Morvan}}, \bibinfo {author} {\bibfnamefont
  {B.}~\bibnamefont {Yoshida}}, \bibinfo {author} {\bibfnamefont
  {N.}~\bibnamefont {Yao}},\ and\ \bibinfo {author} {\bibfnamefont
  {I.}~\bibnamefont {Siddiqi}},\ }\bibfield  {title} {\bibinfo {title} {Quantum
  information scrambling on a superconducting qutrit processor},\ }\href
  {https://doi.org/10.1103/physrevx.11.021010} {\bibfield  {journal} {\bibinfo
  {journal} {Physical Review X}\ }\textbf {\bibinfo {volume} {11}},\ \bibinfo
  {pages} {021010} (\bibinfo {year} {2021})}\BibitemShut {NoStop}%
\bibitem [{\citenamefont {Goss}\ \emph {et~al.}(2022)\citenamefont {Goss},
  \citenamefont {Morvan}, \citenamefont {Marinelli}, \citenamefont {Mitchell},
  \citenamefont {Nguyen}, \citenamefont {Naik}, \citenamefont {Chen},
  \citenamefont {J{\"u}nger}, \citenamefont {Kreikebaum}, \citenamefont
  {Santiago}, \citenamefont {Wallman},\ and\ \citenamefont
  {Siddiqi}}]{Goss_2022}%
  \BibitemOpen
  \bibfield  {author} {\bibinfo {author} {\bibfnamefont {N.}~\bibnamefont
  {Goss}}, \bibinfo {author} {\bibfnamefont {A.}~\bibnamefont {Morvan}},
  \bibinfo {author} {\bibfnamefont {B.}~\bibnamefont {Marinelli}}, \bibinfo
  {author} {\bibfnamefont {B.~K.}\ \bibnamefont {Mitchell}}, \bibinfo {author}
  {\bibfnamefont {L.~B.}\ \bibnamefont {Nguyen}}, \bibinfo {author}
  {\bibfnamefont {R.~K.}\ \bibnamefont {Naik}}, \bibinfo {author}
  {\bibfnamefont {L.}~\bibnamefont {Chen}}, \bibinfo {author} {\bibfnamefont
  {C.}~\bibnamefont {J{\"u}nger}}, \bibinfo {author} {\bibfnamefont {J.~M.}\
  \bibnamefont {Kreikebaum}}, \bibinfo {author} {\bibfnamefont {D.~I.}\
  \bibnamefont {Santiago}}, \bibinfo {author} {\bibfnamefont {J.~J.}\
  \bibnamefont {Wallman}},\ and\ \bibinfo {author} {\bibfnamefont
  {I.}~\bibnamefont {Siddiqi}},\ }\bibfield  {title} {\bibinfo {title}
  {High-fidelity qutrit entangling gates for superconducting circuits},\ }\href
  {https://doi.org/10.1038/s41467-022-34851-z} {\bibfield  {journal} {\bibinfo
  {journal} {Nature Communications}\ }\textbf {\bibinfo {volume} {13}},\
  \bibinfo {pages} {7481} (\bibinfo {year} {2022})}\BibitemShut {NoStop}%
\bibitem [{\citenamefont {Luo}\ \emph {et~al.}(2023)\citenamefont {Luo},
  \citenamefont {Huang}, \citenamefont {Tao}, \citenamefont {Zhang},
  \citenamefont {Zhou}, \citenamefont {Chu}, \citenamefont {Liu}, \citenamefont
  {Wang}, \citenamefont {Cui}, \citenamefont {Liu}, \citenamefont {Yan},
  \citenamefont {Yung}, \citenamefont {Chen}, \citenamefont {Yan},\ and\
  \citenamefont {Yu}}]{Luo_2023}%
  \BibitemOpen
  \bibfield  {author} {\bibinfo {author} {\bibfnamefont {K.}~\bibnamefont
  {Luo}}, \bibinfo {author} {\bibfnamefont {W.}~\bibnamefont {Huang}}, \bibinfo
  {author} {\bibfnamefont {Z.}~\bibnamefont {Tao}}, \bibinfo {author}
  {\bibfnamefont {L.}~\bibnamefont {Zhang}}, \bibinfo {author} {\bibfnamefont
  {Y.}~\bibnamefont {Zhou}}, \bibinfo {author} {\bibfnamefont {J.}~\bibnamefont
  {Chu}}, \bibinfo {author} {\bibfnamefont {W.}~\bibnamefont {Liu}}, \bibinfo
  {author} {\bibfnamefont {B.}~\bibnamefont {Wang}}, \bibinfo {author}
  {\bibfnamefont {J.}~\bibnamefont {Cui}}, \bibinfo {author} {\bibfnamefont
  {S.}~\bibnamefont {Liu}}, \bibinfo {author} {\bibfnamefont {F.}~\bibnamefont
  {Yan}}, \bibinfo {author} {\bibfnamefont {M.-H.}\ \bibnamefont {Yung}},
  \bibinfo {author} {\bibfnamefont {Y.}~\bibnamefont {Chen}}, \bibinfo {author}
  {\bibfnamefont {T.}~\bibnamefont {Yan}},\ and\ \bibinfo {author}
  {\bibfnamefont {D.}~\bibnamefont {Yu}},\ }\bibfield  {title} {\bibinfo
  {title} {Experimental realization of two qutrits gate with tunable coupling
  in superconducting circuits},\ }\href
  {https://doi.org/10.1103/physrevlett.130.030603} {\bibfield  {journal}
  {\bibinfo  {journal} {Physical Review Letters}\ }\textbf {\bibinfo {volume}
  {130}},\ \bibinfo {pages} {030603} (\bibinfo {year} {2023})}\BibitemShut
  {NoStop}%
\bibitem [{\citenamefont {Chen}\ \emph {et~al.}(2025)\citenamefont {Chen},
  \citenamefont {Liu}, \citenamefont {Ma}, \citenamefont {Sun}, \citenamefont
  {Wang}, \citenamefont {Wang}, \citenamefont {Xu}, \citenamefont {Xue},
  \citenamefont {Yan}, \citenamefont {Yang}, \citenamefont {Ding},
  \citenamefont {Gao}, \citenamefont {Li}, \citenamefont {Zhang}, \citenamefont
  {Zhang}, \citenamefont {Jin}, \citenamefont {Yu}, \citenamefont {Chen},\ and\
  \citenamefont {Yan}}]{Chen_2025}%
  \BibitemOpen
  \bibfield  {author} {\bibinfo {author} {\bibfnamefont {Z.}~\bibnamefont
  {Chen}}, \bibinfo {author} {\bibfnamefont {W.}~\bibnamefont {Liu}}, \bibinfo
  {author} {\bibfnamefont {Y.}~\bibnamefont {Ma}}, \bibinfo {author}
  {\bibfnamefont {W.}~\bibnamefont {Sun}}, \bibinfo {author} {\bibfnamefont
  {R.}~\bibnamefont {Wang}}, \bibinfo {author} {\bibfnamefont {H.}~\bibnamefont
  {Wang}}, \bibinfo {author} {\bibfnamefont {H.}~\bibnamefont {Xu}}, \bibinfo
  {author} {\bibfnamefont {G.}~\bibnamefont {Xue}}, \bibinfo {author}
  {\bibfnamefont {H.}~\bibnamefont {Yan}}, \bibinfo {author} {\bibfnamefont
  {Z.}~\bibnamefont {Yang}}, \bibinfo {author} {\bibfnamefont {J.}~\bibnamefont
  {Ding}}, \bibinfo {author} {\bibfnamefont {Y.}~\bibnamefont {Gao}}, \bibinfo
  {author} {\bibfnamefont {F.}~\bibnamefont {Li}}, \bibinfo {author}
  {\bibfnamefont {Y.}~\bibnamefont {Zhang}}, \bibinfo {author} {\bibfnamefont
  {Z.}~\bibnamefont {Zhang}}, \bibinfo {author} {\bibfnamefont
  {Y.}~\bibnamefont {Jin}}, \bibinfo {author} {\bibfnamefont {H.}~\bibnamefont
  {Yu}}, \bibinfo {author} {\bibfnamefont {J.}~\bibnamefont {Chen}},\ and\
  \bibinfo {author} {\bibfnamefont {F.}~\bibnamefont {Yan}},\ }\bibfield
  {title} {\bibinfo {title} {Efficient implementation of arbitrary two-qubit
  gates using unified control},\ }\bibfield  {journal} {\bibinfo  {journal}
  {Nature Physics}\ }\href {https://doi.org/10.1038/s41567-025-02990-x}
  {10.1038/s41567-025-02990-x} (\bibinfo {year} {2025})\BibitemShut {NoStop}%
\bibitem [{\citenamefont {Poyatos}\ \emph {et~al.}(1996)\citenamefont
  {Poyatos}, \citenamefont {Cirac},\ and\ \citenamefont
  {Zoller}}]{Poyatos_1996}%
  \BibitemOpen
  \bibfield  {author} {\bibinfo {author} {\bibfnamefont {J.~F.}\ \bibnamefont
  {Poyatos}}, \bibinfo {author} {\bibfnamefont {J.~I.}\ \bibnamefont {Cirac}},\
  and\ \bibinfo {author} {\bibfnamefont {P.}~\bibnamefont {Zoller}},\
  }\bibfield  {title} {\bibinfo {title} {{Quantum Reservoir Engineering with
  Laser Cooled Trapped Ions}},\ }\href
  {https://doi.org/10.1103/PhysRevLett.77.4728} {\bibfield  {journal} {\bibinfo
   {journal} {Physical Review Letters}\ }\textbf {\bibinfo {volume} {77}},\
  \bibinfo {pages} {4728} (\bibinfo {year} {1996})}\BibitemShut {NoStop}%
\bibitem [{\citenamefont {Diehl}\ \emph {et~al.}(2008)\citenamefont {Diehl},
  \citenamefont {Micheli}, \citenamefont {Kantian}, \citenamefont {Kraus},
  \citenamefont {B\"{u}chler},\ and\ \citenamefont {Zoller}}]{Diehl_2008}%
  \BibitemOpen
  \bibfield  {author} {\bibinfo {author} {\bibfnamefont {S.}~\bibnamefont
  {Diehl}}, \bibinfo {author} {\bibfnamefont {A.}~\bibnamefont {Micheli}},
  \bibinfo {author} {\bibfnamefont {A.}~\bibnamefont {Kantian}}, \bibinfo
  {author} {\bibfnamefont {B.}~\bibnamefont {Kraus}}, \bibinfo {author}
  {\bibfnamefont {H.~P.}\ \bibnamefont {B\"{u}chler}},\ and\ \bibinfo {author}
  {\bibfnamefont {P.}~\bibnamefont {Zoller}},\ }\bibfield  {title} {\bibinfo
  {title} {{Quantum States and Phases in Driven Open Quantum Systems with Cold
  Atoms}},\ }\href {https://doi.org/10.1038/nphys1073} {\bibfield  {journal}
  {\bibinfo  {journal} {Nature Physics}\ }\textbf {\bibinfo {volume} {4}},\
  \bibinfo {pages} {878} (\bibinfo {year} {2008})}\BibitemShut {NoStop}%
\bibitem [{\citenamefont {B\'acsi}\ \emph {et~al.}(2020)\citenamefont
  {B\'acsi}, \citenamefont {Moca},\ and\ \citenamefont {D\'ora}}]{Bacsi_2020}%
  \BibitemOpen
  \bibfield  {author} {\bibinfo {author} {\bibfnamefont {A.}~\bibnamefont
  {B\'acsi}}, \bibinfo {author} {\bibfnamefont {C.~P.}\ \bibnamefont {Moca}},\
  and\ \bibinfo {author} {\bibfnamefont {B.}~\bibnamefont {D\'ora}},\
  }\bibfield  {title} {\bibinfo {title} {{Dissipation-Induced Luttinger Liquid
  Correlations in a One-Dimensional Fermi Gas}},\ }\href
  {https://doi.org/10.1103/PhysRevLett.124.136401} {\bibfield  {journal}
  {\bibinfo  {journal} {Physical Review Letters}\ }\textbf {\bibinfo {volume}
  {124}},\ \bibinfo {pages} {136401} (\bibinfo {year} {2020})}\BibitemShut
  {NoStop}%
\bibitem [{\citenamefont {Wang}\ and\ \citenamefont {Hu}(2020)}]{Wang_2020}%
  \BibitemOpen
  \bibfield  {author} {\bibinfo {author} {\bibfnamefont {Y.-P.}\ \bibnamefont
  {Wang}}\ and\ \bibinfo {author} {\bibfnamefont {C.-M.}\ \bibnamefont {Hu}},\
  }\bibfield  {title} {\bibinfo {title} {{Dissipative Couplings in Cavity
  Magnonics}},\ }\href {https://doi.org/10.1063/1.5144202} {\bibfield
  {journal} {\bibinfo  {journal} {Journal of Applied Physics}\ }\textbf
  {\bibinfo {volume} {127}},\ \bibinfo {pages} {130901} (\bibinfo {year}
  {2020})}\BibitemShut {NoStop}%
\bibitem [{\citenamefont {Garc{\'\i}a-P{\'e}rez}\ \emph
  {et~al.}(2020)\citenamefont {Garc{\'\i}a-P{\'e}rez}, \citenamefont {Rossi},\
  and\ \citenamefont {Maniscalco}}]{GarciaPerez2020IBM}%
  \BibitemOpen
  \bibfield  {author} {\bibinfo {author} {\bibfnamefont {G.}~\bibnamefont
  {Garc{\'\i}a-P{\'e}rez}}, \bibinfo {author} {\bibfnamefont {M.~A.~C.}\
  \bibnamefont {Rossi}},\ and\ \bibinfo {author} {\bibfnamefont
  {S.}~\bibnamefont {Maniscalco}},\ }\bibfield  {title} {\bibinfo {title} {Ibm
  q experience as a versatile experimental testbed for simulating open quantum
  systems},\ }\href {https://doi.org/10.1038/s41534-019-0235-y} {\bibfield
  {journal} {\bibinfo  {journal} {npj Quantum Information}\ }\textbf {\bibinfo
  {volume} {6}},\ \bibinfo {pages} {1} (\bibinfo {year} {2020})}\BibitemShut
  {NoStop}%
\bibitem [{\citenamefont {Ciccarello}\ \emph {et~al.}(2022)\citenamefont
  {Ciccarello}, \citenamefont {Lorenzo}, \citenamefont {Giovannetti},\ and\
  \citenamefont {Palma}}]{Ciccarello2022Collision}%
  \BibitemOpen
  \bibfield  {author} {\bibinfo {author} {\bibfnamefont {F.}~\bibnamefont
  {Ciccarello}}, \bibinfo {author} {\bibfnamefont {S.}~\bibnamefont {Lorenzo}},
  \bibinfo {author} {\bibfnamefont {V.}~\bibnamefont {Giovannetti}},\ and\
  \bibinfo {author} {\bibfnamefont {G.~M.}\ \bibnamefont {Palma}},\ }\bibfield
  {title} {\bibinfo {title} {Quantum collision models: Open system dynamics
  from repeated interactions},\ }\href
  {https://doi.org/10.1016/j.physrep.2022.01.001} {\bibfield  {journal}
  {\bibinfo  {journal} {Physics Reports}\ }\textbf {\bibinfo {volume} {954}},\
  \bibinfo {pages} {1} (\bibinfo {year} {2022})}\BibitemShut {NoStop}%
\bibitem [{\citenamefont {Cao}\ \emph {et~al.}(2023)\citenamefont {Cao},
  \citenamefont {Du},\ and\ \citenamefont {Kou}}]{Cao_2023}%
  \BibitemOpen
  \bibfield  {author} {\bibinfo {author} {\bibfnamefont {K.}~\bibnamefont
  {Cao}}, \bibinfo {author} {\bibfnamefont {Q.}~\bibnamefont {Du}},\ and\
  \bibinfo {author} {\bibfnamefont {S.-P.}\ \bibnamefont {Kou}},\ }\bibfield
  {title} {\bibinfo {title} {{Many-body Non-Hermitian Skin Effect at Finite
  Temperatures}},\ }\href {https://doi.org/10.1103/PhysRevB.108.165420}
  {\bibfield  {journal} {\bibinfo  {journal} {Physical Review B}\ }\textbf
  {\bibinfo {volume} {108}},\ \bibinfo {pages} {165420} (\bibinfo {year}
  {2023})}\BibitemShut {NoStop}%
\bibitem [{FN1()}]{FN1}%
  \BibitemOpen
  \href@noop {} {}\bibinfo {note} {The diagonal and off-diagonal terms of
  $\rho_\text{s}^{(n)}$ for the system qubit in Eq.~\eqref{eq:GKSL} evolve
  independently. It can be shown that the diagonal terms decay more slowly than
  the off-diagonal terms, resulting in inevitable decoherence of the qubit in
  the long term~\cite{Du_2022, Mao_2024}.}\BibitemShut {Stop}%
\bibitem [{\citenamefont {Roberts}\ \emph {et~al.}(2021)\citenamefont
  {Roberts}, \citenamefont {Lingenfelter},\ and\ \citenamefont
  {Clerk}}]{Roberts_2021}%
  \BibitemOpen
  \bibfield  {author} {\bibinfo {author} {\bibfnamefont {D.}~\bibnamefont
  {Roberts}}, \bibinfo {author} {\bibfnamefont {A.}~\bibnamefont
  {Lingenfelter}},\ and\ \bibinfo {author} {\bibfnamefont {A.}~\bibnamefont
  {Clerk}},\ }\bibfield  {title} {\bibinfo {title} {{Hidden Time-Reversal
  Symmetry, Quantum Detailed Balance and Exact Solutions of Driven-Dissipative
  Quantum Systems}},\ }\href {https://doi.org/10.1103/PRXQuantum.2.020336}
  {\bibfield  {journal} {\bibinfo  {journal} {PRX Quantum}\ }\textbf {\bibinfo
  {volume} {2}},\ \bibinfo {pages} {020336} (\bibinfo {year}
  {2021})}\BibitemShut {NoStop}%
\bibitem [{\citenamefont {Chru{\'s}ci{\'n}ski}\ and\ \citenamefont
  {Kossakowski}(2010)}]{Chru_ci_ski_2010}%
  \BibitemOpen
  \bibfield  {author} {\bibinfo {author} {\bibfnamefont {D.}~\bibnamefont
  {Chru{\'s}ci{\'n}ski}}\ and\ \bibinfo {author} {\bibfnamefont
  {A.}~\bibnamefont {Kossakowski}},\ }\bibfield  {title} {\bibinfo {title}
  {Non-markovian quantum dynamics: Local versus nonlocal},\ }\href
  {https://doi.org/10.1103/physrevlett.104.070406} {\bibfield  {journal}
  {\bibinfo  {journal} {Physical Review Letters}\ }\textbf {\bibinfo {volume}
  {104}},\ \bibinfo {pages} {070406} (\bibinfo {year} {2010})}\BibitemShut
  {NoStop}%
\bibitem [{\citenamefont {Hegde}\ \emph {et~al.}(2021)\citenamefont {Hegde},
  \citenamefont {Athulya}, \citenamefont {Pathak}, \citenamefont {Piilo},\ and\
  \citenamefont {Shaji}}]{Hegde_2021}%
  \BibitemOpen
  \bibfield  {author} {\bibinfo {author} {\bibfnamefont {A.~S.}\ \bibnamefont
  {Hegde}}, \bibinfo {author} {\bibfnamefont {K.~P.}\ \bibnamefont {Athulya}},
  \bibinfo {author} {\bibfnamefont {V.}~\bibnamefont {Pathak}}, \bibinfo
  {author} {\bibfnamefont {J.}~\bibnamefont {Piilo}},\ and\ \bibinfo {author}
  {\bibfnamefont {A.}~\bibnamefont {Shaji}},\ }\bibfield  {title} {\bibinfo
  {title} {Open quantum dynamics with singularities: Master equations and
  degree of non-markovianity},\ }\href
  {https://doi.org/10.1103/physreva.104.062403} {\bibfield  {journal} {\bibinfo
   {journal} {Physical Review A}\ }\textbf {\bibinfo {volume} {104}},\ \bibinfo
  {pages} {062403} (\bibinfo {year} {2021})}\BibitemShut {NoStop}%
\bibitem [{SM()}]{SM}%
  \BibitemOpen
  \href@noop {} {}\bibinfo {note} {See Supplemental Material at [URL will be
  inserted by publisher] for the analytical derivation of the quantum master
  equation, detailed calculations for the temporally-correlated dichromatic
  emission, and systematic tests for distinct initial states in the
  LEP-protected quantum synchronization at finite temperatures.}\BibitemShut
  {Stop}%
\bibitem [{\citenamefont {Brown}\ and\ \citenamefont
  {Twiss}(1956)}]{Brown1956}%
  \BibitemOpen
  \bibfield  {author} {\bibinfo {author} {\bibfnamefont {R.~H.}\ \bibnamefont
  {Brown}}\ and\ \bibinfo {author} {\bibfnamefont {R.~Q.}\ \bibnamefont
  {Twiss}},\ }\bibfield  {title} {\bibinfo {title} {Correlation between photons
  in two coherent beams of light},\ }\href {https://doi.org/10.1038/177027a0}
  {\bibfield  {journal} {\bibinfo  {journal} {Nature}\ }\textbf {\bibinfo
  {volume} {177}},\ \bibinfo {pages} {27} (\bibinfo {year} {1956})}\BibitemShut
  {NoStop}%
\bibitem [{\citenamefont {Glauber}(1963)}]{Glauber1963}%
  \BibitemOpen
  \bibfield  {author} {\bibinfo {author} {\bibfnamefont {R.~J.}\ \bibnamefont
  {Glauber}},\ }\bibfield  {title} {\bibinfo {title} {The quantum theory of
  optical coherence},\ }\href {https://doi.org/10.1103/PhysRev.130.2529}
  {\bibfield  {journal} {\bibinfo  {journal} {Physical Review}\ }\textbf
  {\bibinfo {volume} {130}},\ \bibinfo {pages} {2529} (\bibinfo {year}
  {1963})}\BibitemShut {NoStop}%
\bibitem [{\citenamefont {Gatti}\ \emph {et~al.}(2004)\citenamefont {Gatti},
  \citenamefont {Brambilla}, \citenamefont {Bache},\ and\ \citenamefont
  {Lugiato}}]{Gatti2004}%
  \BibitemOpen
  \bibfield  {author} {\bibinfo {author} {\bibfnamefont {A.}~\bibnamefont
  {Gatti}}, \bibinfo {author} {\bibfnamefont {E.}~\bibnamefont {Brambilla}},
  \bibinfo {author} {\bibfnamefont {M.}~\bibnamefont {Bache}},\ and\ \bibinfo
  {author} {\bibfnamefont {L.~A.}\ \bibnamefont {Lugiato}},\ }\bibfield
  {title} {\bibinfo {title} {Ghost imaging with thermal light: Comparing
  entanglement and classical correlation},\ }\href
  {https://doi.org/10.1103/PhysRevLett.93.093602} {\bibfield  {journal}
  {\bibinfo  {journal} {Physical Review Letters}\ }\textbf {\bibinfo {volume}
  {93}},\ \bibinfo {pages} {093602} (\bibinfo {year} {2004})}\BibitemShut
  {NoStop}%
\bibitem [{\citenamefont {Valencia}\ \emph {et~al.}(2005)\citenamefont
  {Valencia}, \citenamefont {Scarcelli}, \citenamefont {D'Angelo},\ and\
  \citenamefont {Shih}}]{Valencia2005}%
  \BibitemOpen
  \bibfield  {author} {\bibinfo {author} {\bibfnamefont {A.}~\bibnamefont
  {Valencia}}, \bibinfo {author} {\bibfnamefont {G.}~\bibnamefont {Scarcelli}},
  \bibinfo {author} {\bibfnamefont {M.}~\bibnamefont {D'Angelo}},\ and\
  \bibinfo {author} {\bibfnamefont {Y.}~\bibnamefont {Shih}},\ }\bibfield
  {title} {\bibinfo {title} {Two-photon imaging with thermal light},\ }\href
  {https://doi.org/10.1103/PhysRevLett.94.063601} {\bibfield  {journal}
  {\bibinfo  {journal} {Physical Review Letters}\ }\textbf {\bibinfo {volume}
  {94}},\ \bibinfo {pages} {063601} (\bibinfo {year} {2005})}\BibitemShut
  {NoStop}%
\bibitem [{\citenamefont {Zhang}\ \emph {et~al.}(2005)\citenamefont {Zhang},
  \citenamefont {Zhai}, \citenamefont {Wu},\ and\ \citenamefont
  {Chen}}]{Zhang2005}%
  \BibitemOpen
  \bibfield  {author} {\bibinfo {author} {\bibfnamefont {D.}~\bibnamefont
  {Zhang}}, \bibinfo {author} {\bibfnamefont {Y.-H.}\ \bibnamefont {Zhai}},
  \bibinfo {author} {\bibfnamefont {L.-A.}\ \bibnamefont {Wu}},\ and\ \bibinfo
  {author} {\bibfnamefont {X.-H.}\ \bibnamefont {Chen}},\ }\bibfield  {title}
  {\bibinfo {title} {Correlated two-photon imaging with true thermal light},\
  }\href {https://doi.org/10.1364/ol.30.002354} {\bibfield  {journal} {\bibinfo
   {journal} {Optics Letters}\ }\textbf {\bibinfo {volume} {30}},\ \bibinfo
  {pages} {2354} (\bibinfo {year} {2005})}\BibitemShut {NoStop}%
\bibitem [{\citenamefont {Tan}\ \emph {et~al.}(2023)\citenamefont {Tan},
  \citenamefont {Yeo}, \citenamefont {Leow}, \citenamefont {Shen},\ and\
  \citenamefont {Kurtsiefer}}]{Tan2023}%
  \BibitemOpen
  \bibfield  {author} {\bibinfo {author} {\bibfnamefont {P.~K.}\ \bibnamefont
  {Tan}}, \bibinfo {author} {\bibfnamefont {X.~J.}\ \bibnamefont {Yeo}},
  \bibinfo {author} {\bibfnamefont {A.~Z.~W.}\ \bibnamefont {Leow}}, \bibinfo
  {author} {\bibfnamefont {L.}~\bibnamefont {Shen}},\ and\ \bibinfo {author}
  {\bibfnamefont {C.}~\bibnamefont {Kurtsiefer}},\ }\bibfield  {title}
  {\bibinfo {title} {Practical range sensing with thermal light},\ }\href
  {https://doi.org/10.1103/PhysRevApplied.20.014060} {\bibfield  {journal}
  {\bibinfo  {journal} {Physical Review Applied}\ }\textbf {\bibinfo {volume}
  {20}},\ \bibinfo {pages} {014060} (\bibinfo {year} {2023})}\BibitemShut
  {NoStop}%
\bibitem [{\citenamefont {Lee}\ \emph {et~al.}(2023)\citenamefont {Lee},
  \citenamefont {Kim}, \citenamefont {Im}, \citenamefont {Kim}, \citenamefont
  {Tamma},\ and\ \citenamefont {Kim}}]{Lee2023}%
  \BibitemOpen
  \bibfield  {author} {\bibinfo {author} {\bibfnamefont {C.-H.}\ \bibnamefont
  {Lee}}, \bibinfo {author} {\bibfnamefont {Y.}~\bibnamefont {Kim}}, \bibinfo
  {author} {\bibfnamefont {D.-G.}\ \bibnamefont {Im}}, \bibinfo {author}
  {\bibfnamefont {U.-S.}\ \bibnamefont {Kim}}, \bibinfo {author} {\bibfnamefont
  {V.}~\bibnamefont {Tamma}},\ and\ \bibinfo {author} {\bibfnamefont {Y.-H.}\
  \bibnamefont {Kim}},\ }\bibfield  {title} {\bibinfo {title} {Coherent
  two-photon lidar with incoherent light},\ }\href
  {https://doi.org/10.1103/PhysRevLett.131.223602} {\bibfield  {journal}
  {\bibinfo  {journal} {Physical Review Letters}\ }\textbf {\bibinfo {volume}
  {131}},\ \bibinfo {pages} {223602} (\bibinfo {year} {2023})}\BibitemShut
  {NoStop}%
\bibitem [{\citenamefont {Burnham}\ and\ \citenamefont
  {Weinberg}(1970)}]{Burnham_1970}%
  \BibitemOpen
  \bibfield  {author} {\bibinfo {author} {\bibfnamefont {D.~C.}\ \bibnamefont
  {Burnham}}\ and\ \bibinfo {author} {\bibfnamefont {D.~L.}\ \bibnamefont
  {Weinberg}},\ }\bibfield  {title} {\bibinfo {title} {Observation of
  simultaneity in parametric production of optical photon pairs},\ }\href
  {https://doi.org/10.1103/physrevlett.25.84} {\bibfield  {journal} {\bibinfo
  {journal} {Physical Review Letters}\ }\textbf {\bibinfo {volume} {25}},\
  \bibinfo {pages} {84} (\bibinfo {year} {1970})}\BibitemShut {NoStop}%
\bibitem [{\citenamefont {Couteau}(2018)}]{Couteau_2018}%
  \BibitemOpen
  \bibfield  {author} {\bibinfo {author} {\bibfnamefont {C.}~\bibnamefont
  {Couteau}},\ }\bibfield  {title} {\bibinfo {title} {{Spontaneous Parametric
  Down-conversion}},\ }\href {https://doi.org/10.1080/00107514.2018.1488463}
  {\bibfield  {journal} {\bibinfo  {journal} {Contemporary Physics}\ }\textbf
  {\bibinfo {volume} {59}},\ \bibinfo {pages} {291} (\bibinfo {year}
  {2018})}\BibitemShut {NoStop}%
\bibitem [{\citenamefont {Alicki}\ \emph {et~al.}(2023)\citenamefont {Alicki},
  \citenamefont {\ifmmode~\check{S}\else \v{S}\fi{}indelka},\ and\
  \citenamefont {Gelbwaser-Klimovsky}}]{Alicki_2023}%
  \BibitemOpen
  \bibfield  {author} {\bibinfo {author} {\bibfnamefont {R.}~\bibnamefont
  {Alicki}}, \bibinfo {author} {\bibfnamefont {M.}~\bibnamefont
  {\ifmmode~\check{S}\else \v{S}\fi{}indelka}},\ and\ \bibinfo {author}
  {\bibfnamefont {D.}~\bibnamefont {Gelbwaser-Klimovsky}},\ }\bibfield  {title}
  {\bibinfo {title} {{Violation of Detailed Balance in Quantum Open Systems}},\
  }\href {https://doi.org/10.1103/PhysRevLett.131.040401} {\bibfield  {journal}
  {\bibinfo  {journal} {Physical Review Letters}\ }\textbf {\bibinfo {volume}
  {131}},\ \bibinfo {pages} {040401} (\bibinfo {year} {2023})}\BibitemShut
  {NoStop}%
\bibitem [{\citenamefont {Pan}\ \emph {et~al.}(2020)\citenamefont {Pan},
  \citenamefont {Chen}, \citenamefont {Chen},\ and\ \citenamefont
  {Zhai}}]{Pan2020}%
  \BibitemOpen
  \bibfield  {author} {\bibinfo {author} {\bibfnamefont {L.}~\bibnamefont
  {Pan}}, \bibinfo {author} {\bibfnamefont {X.}~\bibnamefont {Chen}}, \bibinfo
  {author} {\bibfnamefont {Y.}~\bibnamefont {Chen}},\ and\ \bibinfo {author}
  {\bibfnamefont {H.}~\bibnamefont {Zhai}},\ }\bibfield  {title} {\bibinfo
  {title} {{Non-Hermitian Linear Response Theory}},\ }\href
  {https://doi.org/10.1038/s41567-020-0889-6} {\bibfield  {journal} {\bibinfo
  {journal} {Nature Physics}\ }\textbf {\bibinfo {volume} {16}},\ \bibinfo
  {pages} {767} (\bibinfo {year} {2020})}\BibitemShut {NoStop}%
\bibitem [{\citenamefont {Geier}\ and\ \citenamefont
  {Hauke}(2022)}]{Geier2022}%
  \BibitemOpen
  \bibfield  {author} {\bibinfo {author} {\bibfnamefont {K.~T.}\ \bibnamefont
  {Geier}}\ and\ \bibinfo {author} {\bibfnamefont {P.}~\bibnamefont {Hauke}},\
  }\bibfield  {title} {\bibinfo {title} {{From Non-Hermitian Linear Response to
  Dynamical Correlations and Fluctuation-Dissipation Relations in Quantum
  Many-Body Systems}},\ }\href {https://doi.org/10.1103/PRXQuantum.3.030308}
  {\bibfield  {journal} {\bibinfo  {journal} {PRX Quantum}\ }\textbf {\bibinfo
  {volume} {3}},\ \bibinfo {pages} {030308} (\bibinfo {year}
  {2022})}\BibitemShut {NoStop}%
\bibitem [{\citenamefont {Terashima}\ and\ \citenamefont
  {Ueda}(2005)}]{terashima2005}%
  \BibitemOpen
  \bibfield  {author} {\bibinfo {author} {\bibfnamefont {H.}~\bibnamefont
  {Terashima}}\ and\ \bibinfo {author} {\bibfnamefont {M.}~\bibnamefont
  {Ueda}},\ }\bibfield  {title} {\bibinfo {title} {{Nonunitary Quantum
  Circuit}},\ }\href {https://doi.org/10.1142/S0219749905001456} {\bibfield
  {journal} {\bibinfo  {journal} {International Journal of Quantum
  Information}\ }\textbf {\bibinfo {volume} {03}},\ \bibinfo {pages} {633}
  (\bibinfo {year} {2005})}\BibitemShut {NoStop}%
\bibitem [{\citenamefont {Harrow}\ \emph {et~al.}(2009)\citenamefont {Harrow},
  \citenamefont {Hassidim},\ and\ \citenamefont {Lloyd}}]{Harrow2009}%
  \BibitemOpen
  \bibfield  {author} {\bibinfo {author} {\bibfnamefont {A.~W.}\ \bibnamefont
  {Harrow}}, \bibinfo {author} {\bibfnamefont {A.}~\bibnamefont {Hassidim}},\
  and\ \bibinfo {author} {\bibfnamefont {S.}~\bibnamefont {Lloyd}},\ }\bibfield
   {title} {\bibinfo {title} {{Quantum Algorithm for Linear Systems of
  Equations}},\ }\href {https://doi.org/10.1103/PhysRevLett.103.150502}
  {\bibfield  {journal} {\bibinfo  {journal} {Physical Review Letters}\
  }\textbf {\bibinfo {volume} {103}},\ \bibinfo {pages} {150502} (\bibinfo
  {year} {2009})}\BibitemShut {NoStop}%
\bibitem [{\citenamefont {Uola}\ \emph {et~al.}(2020)\citenamefont {Uola},
  \citenamefont {Costa}, \citenamefont {Nguyen},\ and\ \citenamefont
  {G\"uhne}}]{Uola2020}%
  \BibitemOpen
  \bibfield  {author} {\bibinfo {author} {\bibfnamefont {R.}~\bibnamefont
  {Uola}}, \bibinfo {author} {\bibfnamefont {A.~C.~S.}\ \bibnamefont {Costa}},
  \bibinfo {author} {\bibfnamefont {H.~C.}\ \bibnamefont {Nguyen}},\ and\
  \bibinfo {author} {\bibfnamefont {O.}~\bibnamefont {G\"uhne}},\ }\bibfield
  {title} {\bibinfo {title} {{Quantum Steering}},\ }\href
  {https://doi.org/10.1103/RevModPhys.92.015001} {\bibfield  {journal}
  {\bibinfo  {journal} {Review of Modern Physics}\ }\textbf {\bibinfo {volume}
  {92}},\ \bibinfo {pages} {015001} (\bibinfo {year} {2020})}\BibitemShut
  {NoStop}%
\end{thebibliography}%


\begin{thebibliography}{1}%
\makeatletter
\providecommand \@ifxundefined [1]{%
 \@ifx{#1\undefined}
}%
\providecommand \@ifnum [1]{%
 \ifnum #1\expandafter \@firstoftwo
 \else \expandafter \@secondoftwo
 \fi
}%
\providecommand \@ifx [1]{%
 \ifx #1\expandafter \@firstoftwo
 \else \expandafter \@secondoftwo
 \fi
}%
\providecommand \natexlab [1]{#1}%
\providecommand \enquote  [1]{``#1''}%
\providecommand \bibnamefont  [1]{#1}%
\providecommand \bibfnamefont [1]{#1}%
\providecommand \citenamefont [1]{#1}%
\providecommand \href@noop [0]{\@secondoftwo}%
\providecommand \href [0]{\begingroup \@sanitize@url \@href}%
\providecommand \@href[1]{\@@startlink{#1}\@@href}%
\providecommand \@@href[1]{\endgroup#1\@@endlink}%
\providecommand \@sanitize@url [0]{\catcode `\\12\catcode `\$12\catcode
  `\&12\catcode `\#12\catcode `\^12\catcode `\_12\catcode `\%12\relax}%
\providecommand \@@startlink[1]{}%
\providecommand \@@endlink[0]{}%
\providecommand \url  [0]{\begingroup\@sanitize@url \@url }%
\providecommand \@url [1]{\endgroup\@href {#1}{\urlprefix }}%
\providecommand \urlprefix  [0]{URL }%
\providecommand \Eprint [0]{\href }%
\providecommand \doibase [0]{https://doi.org/}%
\providecommand \selectlanguage [0]{\@gobble}%
\providecommand \bibinfo  [0]{\@secondoftwo}%
\providecommand \bibfield  [0]{\@secondoftwo}%
\providecommand \translation [1]{[#1]}%
\providecommand \BibitemOpen [0]{}%
\providecommand \bibitemStop [0]{}%
\providecommand \bibitemNoStop [0]{.\EOS\space}%
\providecommand \EOS [0]{\spacefactor3000\relax}%
\providecommand \BibitemShut  [1]{\csname bibitem#1\endcsname}%
\let\auto@bib@innerbib\@empty
%</preamble>
\bibitem [{\citenamefont {Jaynes}\ and\ \citenamefont
  {Cummings}(1963)}]{Jaynes1963}%
  \BibitemOpen
  \bibfield  {author} {\bibinfo {author} {\bibfnamefont {E.}~\bibnamefont
  {Jaynes}}\ and\ \bibinfo {author} {\bibfnamefont {F.}~\bibnamefont
  {Cummings}},\ }\bibfield  {title} {\bibinfo {title} {Comparison of quantum
  and semiclassical radiation theories with application to the beam maser},\
  }\href {https://doi.org/10.1109/proc.1963.1664} {\bibfield  {journal}
  {\bibinfo  {journal} {Proceedings of the IEEE}\ }\textbf {\bibinfo {volume}
  {51}},\ \bibinfo {pages} {89} (\bibinfo {year} {1963})}\BibitemShut {NoStop}%
\end{thebibliography}%

\title{End Matter}

\maketitle

\renewcommand{\theequation}{A\arabic{equation}}
\setcounter{equation}{0}

\section{End Matter A: Quantum master equation}

At the $n$th collision step in the quantum circuits [Fig.~\ref{fig:fig1}(c)], the non-interacting Hamiltonian is given by $H^{(n)}_0 = H_\text{s} + H^{(n)}_\text{q}$, consisting of the system Hamiltonian $H_\text{s}$, and the reservoir qubit Hamiltonian $H^{(n)}_\text{q}$ involved in this collision.
The \textit{bare} density matrices $\rho^{(n)}$ and $\rho^{(n+1)}$, representing the states at the start and end of this collision in the Schr$\ddot{\text{o}}$dinger picture, can be transformed into their counterparts $\rho^{(n)\text{I}}$ and $\rho^{(n+1)\text{I}}$ in the interaction picture, respectively.
These transformations are given by $\rho^{(n)\text{I}} = e^{i H^{(n)}_0 \bar{t}} \rho^{(n)} e^{-i H^{(n)\dag}_0 \bar{t}}$ and $\rho^{(n+1)\text{I}} = e^{i H^{(n)}_0 \bar{t}} \rho^{(n+1)} e^{-i H^{(n)\dag}_0 \bar{t}}$, where ``I" highlights the interaction picture.
Similarly, the operators transform as follows: $A^{(n)\text{I}} =e^{-i H_\text{s} \bar{t}} A^{(n)} e^{i H_\text{s} \bar{t}}$, $B^{(n)\text{I}} = e^{-i H^{(n)}_\text{q} \bar{t}} B^{(n)} e^{i H^{(n)}_\text{q} \bar{t}}$, and $H^{(n)\text{I}}_\text{sq} = e^{-i H^{(n)}_0 \bar{t}} H^{(n)}_\text{sq} e^{i H^{(n)}_0 \bar{t}}$.

At the beginning of the $n$th collision step, $\rho^{(n)\text{I}}$ is the product of the density matrix $\rho_\text{s}^{(n)\text{I}}$ for the system and the density matrix $\rho^{(n)}_\text{q}$ for the involved qubit, i.e., $\rho^{(n)\text{I}} = \rho_\text{s}^{(n)\text{I}} \otimes \rho^{(n)}_\text{q}$. The former $\rho^{(n)\text{I}}_\text{s}$ is obtained by taking a partial trace over the non-orthogonal bases for the reservoir qubit at the end of the last collision step $n-1$.
The latter is prepared using quantum circuit techniques~\cite{Chen_2025}.
It is important to note that $\rho^{(n)\text{I}}_\text{q} \equiv \rho^{(n)}_\text{q}$, following the convention defined in the main text.

For the weak collision $g\bar{t} \ll 1$, the time-evolution operator $U^{(n)\text{I}}_\text{sq} = e^{-i g H^{(n)\text{I}}_\text{sq} \bar{t}}$ in the interaction picture can be expanded to second order as
\begin{eqnarray}
U^{(n)\text{I}}_\text{sq} \approx \mathbbm{1} - i g\bar{t} H^{(n)\text{I}}_\text{sq} - \frac{g^2\bar{t}^2}{2} \left(H^{(n)\text{I}}_\text{sq} \right)^2\, .
\end{eqnarray}
Thus the difference is given by
\begin{eqnarray}\label{eq:2ndorderBMME}
\begin{split}
& \frac{\rho_\text{s}^{(n+1)\text{I}} - \rho_\text{s}^{(n)\text{I}}}{\bar{t}} = \\
& g^2\bar{t}\ \text{tr}_\text{q} \! \left[ H^{(n)\text{I}}_\text{sq} \rho^{(n)\text{I}} H_\text{sq}^{(n)\text{I}\dag}\! - \! \frac{1}{2}\left\{\left(H^{(n)\text{I}}_\text{sq} \right)^2,\, \rho^{(n)\text{I}}\right\}_\dag \right]\, ,
\end{split}
\end{eqnarray}
where we use the stability condition $\text{tr}_{\text{q}} [ B^{(n)}_{\phantom{\text{q}}} \rho^{(n)}_{\text{q}} ]=0$ to remove the first-order coherent drift term, which is formally analogous to a Lamb-shift Hamiltonian renormalization in standard QME derivations~\cite{Breuer_2007, Du_2022}.
We note that the stability condition can always be enforced by redefining $B^{(n)} = B^{'(n)} - \mu_\text{b}$ if $B^{'(n)}$ gives $\text{tr}_\text{q} [B^{'(n)}\rho^{(n)}_\text{q}]=\mu_\text{b}\ne 0$.

Further, under the weak-coupling condition $g^2\bar{t}\ll \omega$ discussed in the main text, we retain only the resonant part $H^{(n)}_\text{sq} \approx \sum_{\omega}A^{(n)}_\omega \otimes B^{(n)}_{-\omega}$, where $\omega$ takes the values $\pm \omega_l$.
This procedure is equivalent to directly invoking the rotating-wave approximation.
Thus, Eq.~\eqref{eq:2ndorderBMME} simplifies to
\begin{eqnarray}\label{eq:QMEI}
\begin{split}
& \frac{\rho_\text{s}^{(n+1)\text{I}} - \rho_\text{s}^{(n)\text{I}}}{\bar{t}} = g^2\bar{t}\sum_{\omega=\pm \omega_{l}} \left(\bar{\gamma}^{(n)}_\omega A^{(n)}_\omega \rho_\text{s}^{(n)\text{I}} A^{(n)\dag}_\omega\right.\\
& \left. - \frac{1}{2} \left\{\gamma^{(n)}_\omega A^{(n)}_{-\omega}A^{(n)}_\omega,\, \rho_\text{s}^{(n)\text{I}} \right\}_\dag\right)\, ,
\end{split}
\end{eqnarray}
where the dual spectral functions $\gamma^{(n)}_{\omega}$ and $\bar{\gamma}^{(n)}_{\omega}$ follow the definitions given in Eq.~\eqref{eq:SpectralFunctions} of the main text.
Finally, returning to the Schr$\ddot{\text{o}}$dinger picture, Eq.~\eqref{eq:QMEI} yields QME in Eq.~\eqref{eq:GKSL} of the main text.

\vspace{0.3cm}

\section{End Matter B: Two inequivalent spectral functions}

We consider the transition from level ``$a$" to level ``$b$" as an example, where $\omega = \omega_l$. In that case, we have
\begin{eqnarray}\label{eq:collision_spectralfunc}
\begin{split}
\gamma^{(n)}_{\omega_l} & = \sum_{\alpha = a,\,b} w_\alpha^l \bra{\alpha_\text{R}} B^{(n)}_{\omega_l} B^{(n)}_{-\omega_l} \ket{\alpha_\text{R}}\\
&= w_b^l \bra{b_\text{L}} B^{(n)} \ket{a_\text{R}}\bra{a_\text{L}} B^{(n)} \ket{b_\text{R}}\\
& =w_b^l \mathbbm{B}_{ba}^{(n)} \mathbbm{B}_{ab}^{(n)}\, ,\\
\bar{\gamma}^{(n)}_{\omega_l} &= \sum_{\alpha = a,\,b} w_\alpha^l \bra{\alpha_\text{R}} B^{(n)\dag}_{-\omega_l} B^{(n)}_{-\omega_l} \ket{\alpha_\text{R}}\\
&= w_b^l \bra{b_\text{R}} B^{(n)\dag} \ket{a_\text{L}} \bra{a_\text{L}} B^{(n)} \ket{b_\text{R}}\\
& =w_b^l \mathbbm{B}_{ab}^{(n)*} \mathbbm{B}_{ab}^{(n)} \, .
\end{split}
\end{eqnarray}
In the above derivation, we use $\mathbbm{B}^{(n)}_{ab} = \bra{a_\text{L}}B^{(n)} \ket{b_\text{R}}$, $\mathbbm{B}^{(n)}_{ba} = \bra{b_\text{L}} B^{(n)} \ket{a_\text{R}}$, along with the normalization condition $\braket{a_\text{R} \vert a_\text{R}} =  \braket{b_\text{R} \vert b_\text{R}} = 1$ for the biorthonormal left eigenstates $\bra{a_\text{L}}$, $\bra{b_\text{L}}$ and right eigenstates $\ket{a_\text{R}}$, $\ket{b_\text{R}}$ of the reservoir qubits.

For conventional thermal reservoirs, where $\ket{a_\text{L}} = \ket{a_\text{R}}$ and $\ket{b_\text{L}} = \ket{b_\text{R}}$, the biorthonormalization condition between left and right eigenstates reduces to the orthonormalization condition.
In this case, it is clear that $\gamma^{(n)}_{\omega_l} = \bar{\gamma}^{(n)}_{\omega_l}$, since $\mathbbm{B}^{(n)}_{ab} = \mathbbm{B}^{(n)*}_{ba}$ for the Hermitian operator $B^{(n)}$.
However, for two applications of this platform discussed in the main text, the right eigenstates are not equal to the left ones, i.e., $\ket{a_\text{L}} \ne \ket{a_\text{R}}$ and $\ket{b_\text{L}} \ne \ket{b_\text{R}}$.
Consequently, the orthogonality of the right eigenstates, i.e., $\braket{a_\text{R} \vert b_\text{R}}=0$, does not hold when we maintain $\braket{a_\text{L} \vert b_\text{R}}=0$ in the biorthonormalization condition.
Therefore, $\gamma^{(n)}_{\omega_l}$ differs from $\bar{\gamma}^{(n)}_{\omega_l}$ when $\mathbbm{B}^{(n)}_{ab} \ne \mathbbm{B}^{(n)*}_{ba}$.
This analysis also applies to the case where $\omega=-\omega_l$.

\end{document}

\title{\textbf{\large Supplemental Material for:\\ Encoding complex-balanced thermalization in quantum circuits}}

\author{Yiting Mao}
\altaffiliation{These authors contributed equally to this work.}
\affiliation{School of Physics, Zhejiang University, Hangzhou, 310058, China}
\affiliation{Beijing Computational Science Research Center, Beijing 100084, China}

\author{Peigeng Zhong}
\altaffiliation{These authors contributed equally to this work.}
\affiliation{School of Physics, Harbin Institute of Technology, Harbin 150001, China}

 \author{Haiqing Lin}
 \email[]{haiqing0@csrc.ac.cn}
 \affiliation{School of Physics, Zhejiang University, Hangzhou, 310058, China}
\affiliation{Institute for Advanced Studies of Physics, Zhejiang University, Hangzhou, 310058, China}

 \author{Xiaoqun Wang}
 \email[]{xiaoqunwang@zju.edu.cn}
 \affiliation{School of Physics, Zhejiang University, Hangzhou, 310058, China}
\affiliation{Institute for Advanced Studies of Physics, Zhejiang University, Hangzhou, 310058, China}

\author{Shijie Hu}
\email[]{shijiehu@csrc.ac.cn}
\affiliation{Beijing Computational Science Research Center, Beijing 100084, China}
\affiliation{Department of Physics, Beijing Normal University, Beijing, 100875, China}

\maketitle

\section{Quantum master equation}
In this section, we provide further details regarding the analytical derivation of the quantum master equation (QME).
Following the convention defined in the main text and End Matter, for the weak collision $g\bar{t}\ll 1$, the time-evolution operator $U_\text{sq}^{(n)\text{I}}=e^{-igH_\text{sq}^{(n)\text{I}}\bar{t}}$ in the interaction picture can be expanded to second order as 
\begin{eqnarray}
U^{(n)\text{I}}_\text{sq} \approx \mathbbm{1} - i g\bar{t} H^{(n)\text{I}}_\text{sq} - \frac{g^2\bar{t}^2}{2} \left(H^{(n)\text{I}}_\text{sq} \right)^2\, .
\end{eqnarray}
Thus, we have 
\begin{align}\label{eq:CollisionStep}
\rho_\text{s}^{(n+1)\text{I}}&=\text{tr}_\text{q}\left[U_\text{sq}^{(n)\text{I}}\left(\rho_\text{s}^{(n)\text{I}}\otimes \rho_\text{q}^{(n)}\right)U_\text{sq}^{(n)\text{I}\dag}\right]\nonumber\\
&=\text{tr}_\text{q}\left[\left(\mathbbm{1} - i g\bar{t} H^{(n)\text{I}}_\text{sq} - \frac{g^2\bar{t}^2}{2} \left(H^{(n)\text{I}}_\text{sq} \right)^2\right)\left(\rho_\text{s}^{(n)\text{I}}\otimes \rho_\text{q}^{(n)}\right)\left(\mathbbm{1} + i g\bar{t} H^{(n)\text{I}\dag}_\text{sq} - \frac{g^2\bar{t}^2}{2} \left(H^{(n)\text{I}\dag}_\text{sq} \right)^2\right)\right]\nonumber\\
&\approx \text{tr}_\text{q}\left[\left(\rho_\text{s}^{(n)\text{I}} \otimes \rho_\text{q}^{(n)}\right) - i g\bar{t} \left[H_\text{sq}^{(n)\text{I}},\,\left(\rho_\text{s}^{(n)\text{I}}\otimes \rho_\text{q}^{(n)}\right) \right]_\dag+g^2\bar{t}^2 H_\text{sq}^{(n)\text{I}}\left(\rho_\text{s}^{(n)\text{I}} \otimes \rho_\text{q}^{(n)}\right) H_\text{sq}^{(n)\text{I}\dag}\right.\nonumber\\
& \quad \left.-\frac{g^2\bar{t}^2}{2}\left\{\left(H_\text{sq}^{(n)\text{I}}\right)^2,\,\left(\rho_\text{s}^{(n)\text{I}}\otimes \rho_\text{q}^{(n)}\right)\right\}_\dag\right] + \mathcal{O}[(g\bar{t})^3]\, ,
\end{align}
where $[O_1,O_2]_{\dagger}\equiv O_1O_2-O_2O_1^\dagger$ and $\{O_1,O_2\}_{\dagger}\equiv O_1 O_2 + O_2 O^{\dagger}_1$ denote a generalized commutator and anticommutator between operators $O_1$ and $O_2$, respectively.
Next, we neglect $\mathcal{O}[(g\bar{t})^3]$ terms. The first-order term is proportional to $\text{tr}_\text{q}\left[B^{(n)}\rho_\text{q}^{(n)}\right]$ and its adjoint. We impose the stability condition $\text{tr}_\text{q}\left[B^{(n)}\rho_\text{q}^{(n)}\right]=0$, which eliminates this drift term.
Thus we arrive at the following expression
\begin{align}\label{eq:CollisionStepN}
\rho_\text{s}^{(n+1)\text{I}}&=\rho_\text{s}^{(n)\text{I}}+g^2\bar{t}^2\ \text{tr}_\text{q}\left[H_\text{sq}^{(n)\text{I}}\left(\rho_\text{s}^{(n)\text{I}}\otimes \rho_\text{q}^{(n)}\right)H_\text{sq}^{(n)\text{I}\dag}-\frac{1}{2}\left\{\left(H_\text{sq}^{(n)\text{I}}\right)^2,\,\left(\rho_\text{s}^{(n)\text{I}}\otimes \rho_\text{q}^{(n)}\right)\right\}_\dag\right]\, ,
\end{align}
which gives rise to Eq.~(A2) in End Matter~A.

For long-term dynamics, we retain only the resonant part $H^{(n)}_\text{sq} \approx \sum_{\omega}A^{(n)}_\omega \otimes B^{(n)}_{-\omega}$, as discussed in the main text. Thus, Eq.~\eqref{eq:CollisionStepN} simplifies to
\begin{eqnarray}
\begin{split}
\Delta \rho_\text{s}^{(n)\text{I}} &=g^2\bar{t}\sum_\omega \text{tr}_\text{q}\left[\left(A_\omega^{(n)}\otimes B_{-\omega}^{(n)}\right)\left(\rho_\text{s}^{(n)\text{I}}\otimes \rho_\text{q}^{(n)}\right) \left(A_\omega^{(n)\dag}\otimes B_{-\omega}^{(n)\dag}\right)\right.\\
&\quad \left.-\frac{1}{2}\left(A_{\omega}^{(n)}\otimes B_{-\omega}^{(n)}\right)\left(A_{-\omega}^{(n)}\otimes B_{\omega}^{(n)}\right)\left(\rho_\text{s}^{(n)\text{I}}\otimes \rho_\text{q}^{(n)}\right)-\frac{1}{2}\left(\rho_\text{s}^{(n)\text{I}}\otimes \rho_\text{q}^{(n)}\right)\left(A_{-\omega}^{(n)\dag}\otimes B_{\omega}^{(n)\dag}\right)\left(A_{\omega}^{(n)\dag}\otimes B_{-\omega}^{(n)\dag}\right)\right]\\
&=g^2\bar{t}\sum_\omega \left( \text{tr}_\text{q}\left[ B_{-\omega}^{(n)}\rho_\text{q}^{(n)}B_{-\omega}^{(n)\dag}\right] A_\omega^{(n)}\rho_\text{s}^{(n)\text{I}}A_\omega^{(n)\dag}\right.\\
&\quad \left.-\frac{1}{2}\text{tr}_\text{q} \left[B_{-\omega}^{(n)}B_{\omega}^{(n)}\rho_\text{q}^{(n)}\right] A_\omega^{(n)}A_{-\omega}^{(n)}\rho_\text{s}^{(n)\text{I}}-\frac{1}{2}\text{tr}_\text{q}\left[\rho_\text{q}^{(n)}B_\omega^{(n)\dag}B_{-\omega}^{(n)\dag}\right]\rho_\text{s}^{(n)\text{I}}A_{-\omega}^{(n)\dag}A_\omega^{(n)\dag}\right)\\
&=g^2\bar{t}\sum_\omega \left(\bar{\gamma}^{(n)}_\omega A^{(n)}_\omega \rho_\text{s}^{(n)\text{I}} A^{(n)\dag}_\omega - \frac{1}{2} \left\{\gamma^{(n)}_\omega A^{(n)}_{-\omega}A^{(n)}_\omega,\, \rho_\text{s}^{(n)\text{I}} \right\}_\dag\right)\, ,
\end{split}
\end{eqnarray}
which yields Eq.~(A3) in End Matter~A. Note that the prefactor $g^2\bar{t}$ governs the rate at which the system approaches complex-balanced thermalization, and is thus referred to as the \textit{dissipation rate}.

\section{Temporally correlated dichromatic emission}

We provide the details of the calculation for the temporally correlated dichromatic emission discussed in the main text. 
At collision step $n$, the full Hamiltonian is given by
\begin{eqnarray}\label{eq:H_total}
% H_\text{total} = H_\text{s}+H_\text{q}^{(n)}+\tcr{g}H_\text{sq}^{(n)}+H_\text{p}+\tcr{g_\text{int}}H_\text{sp}\, ,
H_\text{total}^{(n)} = H_\text{s}+H_\text{p}+g_\text{int}H_\text{sp}+H_\text{q}^{(n)}+gH_\text{sq}^{(n)}\, ,
\end{eqnarray}
where $H_\text{s}$ is the Hamiltonian for the three-level system (with energy levels $\ket{0}$, $\ket{1}$, and $\ket{2}$), and $H_\text{p}=\omega_{21} p_1^\dag p_1+\omega_{10} p_2^\dag p_2$ represents the Hamiltonian of the two free photonic (p) modes $p_1$ and $p_2$.
% The system interacts with these two modes via the Jaynes-Cummings term $H_\text{sp} =\ketbra{2}{1} p_1 + \ketbra{1}{0} p_2 + \textrm{h.c.}$, where $g_\text{int}$ denotes the interaction strength in $H_\text{total}$.
The system interacts with these two photonic modes via the Jaynes-Cummings term $H_\text{sp} =\ketbra{2}{1} p_1 + \ketbra{1}{0} p_2 + \textrm{h.c.}$, where $g_\text{int}$ denotes the interaction strength in $H_\text{total}^{(n)}$. We also consider $H_\text{sq}^{(n)}=A^{(n)}\otimes B^{(n)}$.
This Jaynes-Cummings form corresponds to the rotating-wave approximation, with each photonic mode resonantly coupled to its corresponding transition in the system~\cite{Jaynes1963}.
In the dilute-photon limit, we restrict the maximum photon number to $N^\text{max}_\text{p}=2$ for each mode. This cutoff captures the relevant two-photon coincidences for $G^{(2)}$ while higher-photon sectors remain negligibly occupied.

\begin{figure}[b]
\begin{center}
\includegraphics[width=0.83\columnwidth]{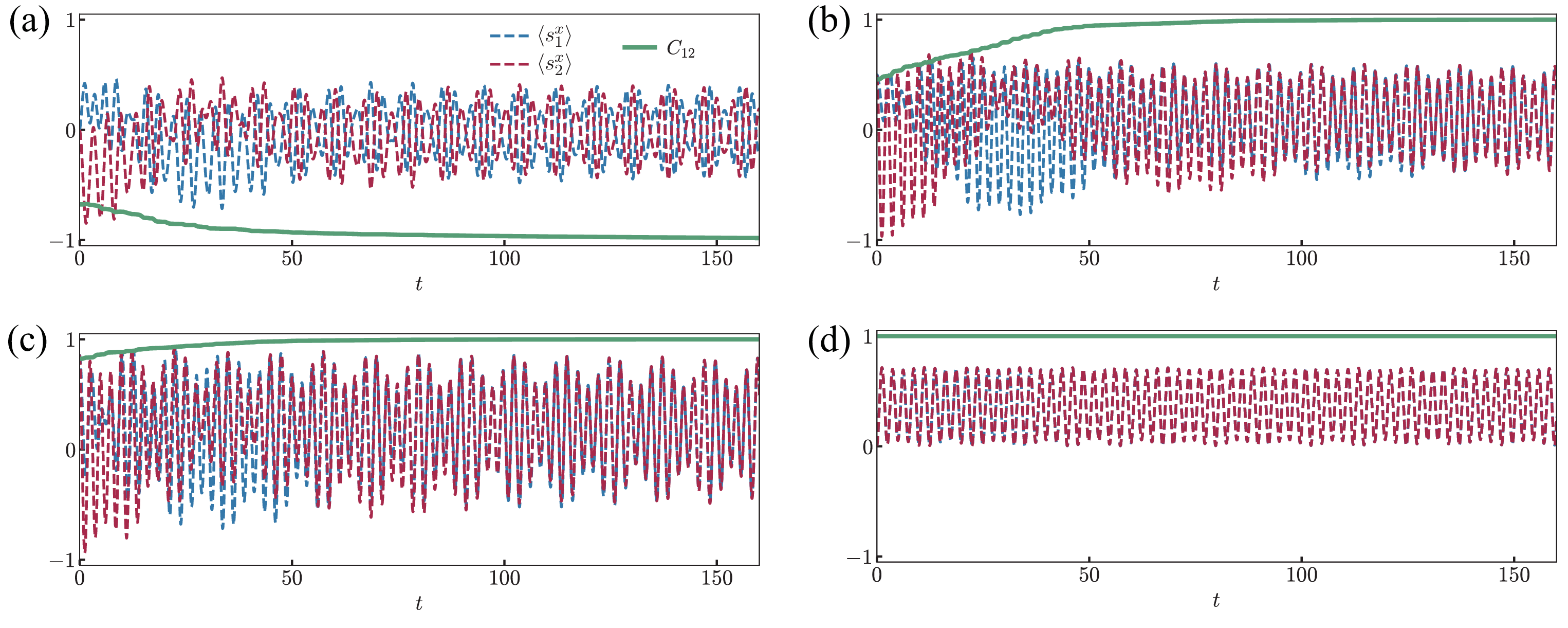}
\caption{\label{fig:SMFig1}
(a) Long-term perfect LEP-protected anti-phase QS with $C_{12}=-1$ for a ferromagnetic Ising-type interaction with $J<0$.
(b)-(d) LEP-protected QS is established, starting from different initial states $\rho_\text{s}^{(0)}$, with the relative angle between spins $\mathbf{s}_1$ and $\mathbf{s}_2$ set to (b) $2\pi/3$, (c) $\pi/3$ and (d) $0$.
Other parameters are the same as those used in Fig.~4(d) of the main text.
}
\end{center}
\end{figure}

To demonstrate photon correlations after the three-level system and photon fields reach complex-balanced thermalization, we monitor the joint system-field density matrix $\rho_\text{sp}^{(n)}$ during time evolution.
The initial state is chosen as $\rho_\text{sp}^{(0)} = \ketbra{0}{0} \otimes \rho_\text{p}^{(0)}$, and $\rho_\text{p}^{(0)}$ is prepared in the photon vacuum state. The dynamics of $\rho_\text{sp}^{(n)}$ is governed by the equation
\begin{eqnarray}\label{eq:dynamics}
\Delta \rho_\text{sp}^{(n)} = \frac{\rho_\text{sp}^{(n+1)} - \rho_\text{sp}^{(n)}}{\bar{t}} &=&-i\left[H_\text{s}+H_\text{p}+g_\text{int}H_\text{sp},\,\rho_\text{sp}^{(n)}\right] +\mathcal{L}_\text{j}^{\text{s}}[\rho_\text{sp}^{(n)}]-\mathcal{L}_\text{d}^{\text{s}}[\rho_\text{sp}^{(n)}]+\mathcal{D}[\rho_\text{sp}^{(n)}]\, .
\end{eqnarray}
% \Delta \rho_\text{sp}^{(n)} = \frac{\rho_\text{sp}^{(n+1)} - \rho_\text{sp}^{(n)}}{\bar{t}} = \text{tr}_\text{q}\left[\mathcal{L}[\rho_\text{sp}^{(n)}] \right] + \mathcal{D} [\rho_\text{sp}^{(n)} ]\, ,
% where the Liouvillian superoperator $\mathcal{L}$ is given by
% \begin{eqnarray}
%     \mathcal{L} [\rho_\text{sp}^{(n)}] = -i\left[H_{\rm total},\,\rho_\text{sp}^{(n)}\otimes \rho_\text{q}^{(n)}\right]_\dag
% \end{eqnarray}
% under the collision map in Eq.~(1) of the main text.
% It becomes
% \begin{eqnarray}
%     \mathcal{L} [\rho_\text{sp}^{(n)}] = -i\left[H_\text{s}+H_\text{p}+H_\text{sp},\,\rho_\text{sp}^{(n)}\right]+\left(\mathcal{L}_\text{j} [\rho_\text{s}^{(n)}]-\mathcal{L}_\text{d} [\rho_\text{s}^{(n)}]\right)\otimes \mathbbm{1}_\text{p}
% \end{eqnarray}
% in the QME defined by Eq.~(2) of the main text.
% Thus, the density matrix of the system is defined as $\rho_\text{s}^{(n)}=\text{tr}_\text{p} \left[\rho_\text{sp}^{(n)}\right]$, where $\text{tr}_\text{p} [\cdots]$ denotes the partial trace over the photonic degrees of freedom.
Here $\mathcal{L}_{\text{j}}^{\text{s}}$ and $\mathcal{L}_{\text{d}}^{\text{s}}$ are derived from Eq.~(2) of the main text by the substitutions $A_\omega^{(n)}\rightarrow A_\omega^{(n)}\otimes \mathbbm{1}_\text{p}$ and $\rho_\text{s}^{(n)}\rightarrow\rho_\text{sp}^{(n)}$.
Numerical simulation data for the collision model are obtained by the time evolution defined in Eq.~(1) of the main text, where the emitter and photonic modes are treated as a composite system described by the full Hamiltonian $H_\text{total}^{(n)}$.
For the steady-state curves in Fig.~3(c) of the main text, we evaluate
\begin{eqnarray}
G^{(2)}_{x_1x_2}(n')
=
\frac{\braket{p^\dag_{x_1}p_{x_2}^{(n')\dag}p_{x_2}^{(n')}p_{x_1}}_n}
{\braket{p^\dag_{x_1}p_{x_1}}_n\braket{p^\dag_{x_2}p_{x_2}}_n},
\end{eqnarray}
with the expectation value taken after $\rho_\text{sp}^{(n)}$ reaches the steady state.
To account for the dissipation of the photonic modes, we introduce additional Lindblad terms $L_x = \sqrt{\kappa} p_x$ ($x=1$, $2$) with a dissipation rate $\kappa$, leading to 
\begin{eqnarray}
    \mathcal{D} [\rho_\text{sp}^{(n)}] = \sum_{x=1,2} \kappa\left[\left(\mathbbm{1}_\text{s}\otimes p_x\right)\rho_\text{sp}^{(n)}\left(\mathbbm{1}_\text{s}\otimes p_x^\dag\right) -\frac{1}{2}\left\{\mathbbm{1}_\text{s}\otimes p_x^\dag p^{\phantom{\dag}}_x,\,\rho_\text{sp}^{(n)}\right\}\right]\, .
\end{eqnarray}
Figure~3(b) of the main text compares the time evolution of photon numbers obtained from the collision map and QME.
When $g\bar{t}\ll 1$, the two methods show excellent agreement.

%\begin{figure}[t]
%\begin{center}
%\includegraphics[width=0.83\columnwidth]{SMFig2.eps}
%\caption{\label{fig:SMFig2}
%Time evolution of the expectation value $\braket{s^x_1}$ for spin $\mathbf{s}_1$, obtained from the collision map in Eq.~(1) and QME in Eq.~(2) of the main text.
%\tcr{In panels (a)-(c), we vary $\bar{t}=0.2$, $0.05$, and $0.005$ at fixed $g=1$, corresponding to $g\bar{t}=0.2$, $0.05$, and $0.005$, respectively; in panels (d)-(f), we vary $g=0.5$, $0.1$, and $0.01$ at fixed $\bar{t}=0.2$, corresponding to $g\bar{t}=0.1$, $0.02$, and $0.002$, respectively.}
%\tcr{Other parameters are the same as those used in Fig.~4(d) of the main text, except for the displayed values of $g$ and $\bar{t}$; the reference main-text choice is $g=2$ and $\bar{t}=0.05$.}
%}
%\end{center}
%\end{figure}

\section{LEP-protected Quantum synchronization at finite temperatures}

We present additional numerical results on Liouvillian-exceptional-point (LEP) protected quantum synchronization (QS) at finite temperatures.
Figure~\ref{fig:SMFig1}(a) demonstrates perfect anti-phase QS ($C_{12}=-1$) for the ferromagnetic Ising-type interaction with $J<0$. In addition, Figs.~\ref{fig:SMFig1}(b)-\ref{fig:SMFig1}(d) show benchmarks using different initial states $\rho_\text{s}^{(0)}$, with perfect in-phase QS consistently established in the long term.
%Finally, we explore \tcr{the distinct effects of varying either the collision interval $\bar{t}$ or the coupling strength $g$, while tracking the corresponding weak-collision parameter $g\bar{t}$}, and examine the discrepancies in the dynamics described by the collision map in Eq.~(1) and QME in Eq.~(2) of the main text.
%\tcr{From Figs.~\ref{fig:SMFig2}(a)-(c), we observe that when QS is present, i.e., when the Liouvillian modes with maximal real part $\mathrm{Re}\lambda=0$ carry nonzero imaginary parts, decreasing $\bar{t}$ from $0.2$ to $0.005$ at fixed $g=1$ (or decreasing $g\bar{t}$ from $0.2$ to $0.005$) does not improve the agreement between the dynamics obtained from the two methods.}
%The reason is that, in this regime, the Liouvillian gap closes, and time periods introduced by the two oscillation modes are no longer governed by the dissipation rate $g^2\bar{t}$~\cite{Filippov2022}.
%This indicates that the dynamics deviate from the standard Born-Markov approximation~\cite{Luchnikov2017, Filippov2020}.
%\tcr{However, when $g$ is reduced from $0.5$ to $0.01$ at fixed $\bar{t}=0.2$, i.e., decreasing $g\bar{t}$ from $0.1$ to $0.002$ through weaker coupling [Figs.~\ref{fig:SMFig2}(d)-(f)], the dynamics obtained from the two methods align well because higher-order coherent corrections are suppressed toward the von-Hove weak-coupling limit.}

\bibliography{sm_refs}